\documentclass[a4paper,fleqn]{cas-sc}

\usepackage[authoryear,longnamesfirst]{natbib}
\usepackage{graphicx}
\usepackage{multirow}
\usepackage{natbib}
\usepackage{xspace}
\usepackage{subfig}
\usepackage{amsmath}
\usepackage{caption}
\usepackage{xspace}
\newcommand{\toola}{PeerWise\xspace}

\newcommand{\toolb}{RiPPLE\xspace}

\def\tsc#1{\csdef{#1}{\textsc{\lowercase{#1}}\xspace}}
\tsc{WGM}
\tsc{QE}
\tsc{EP}
\tsc{PMS}
\tsc{BEC}
\tsc{DE}

\begin{document}
\let\WriteBookmarks\relax
\def\floatpagepagefraction{1}
\def\textpagefraction{.001}
\shorttitle{Learnersourcing in the Age of AI}
\shortauthors{Khosravi et al.}


{
\title [mode = title]{Learnersourcing in the Age of AI: Student, Educator and Machine Partnerships for Content Creation}









\author[1]{Hassan Khosravi}[%
     orcid=0000-0001-8664-6117,
    ]
\ead{h.khosravi@uq.edu.au}
\ead[URL]{http://hassan-khosravi.net/}
\address[1]{The University of Queensland, Queensland, Australia}

\author[2]{Paul Denny}[%
     orcid=0000-0002-5150-9806,
  ]
\ead{paul@cs.auckland.ac.nz}
\ead[URL]{https://profiles.auckland.ac.nz/p-denny}
\address[2]{The University of Auckland, Auckland, New Zealand}

\author[3]{Steven Moore}[%
     orcid=0000-0002-5256-0339,
  ]
\ead{StevenMo@andrew.cmu.edu}
\ead[URL]{https://www.hcii.cmu.edu/people/steven-moore}
\address[3]{Carnegie Mellon University, Pittsburgh, United States}

\author[3]{John Stamper}[%
     orcid=0000-0002-2291-1468,
  ]
\ead{jstamper@cs.cmu.edu}
\ead[URL]{https://www.hcii.cmu.edu/people/john-stamper}













\begin{abstract}
Engaging students in creating novel content, also referred to as learnersourcing, is increasingly recognised as an effective approach to promoting higher-order learning, deeply engaging students with course material and developing large repositories of content suitable for personalized learning. Despite these benefits, some common concerns and criticisms are associated with learnersourcing (e.g., the quality of resources created by students, challenges in incentivising engagement and lack of availability of reliable learnersourcing systems), which have limited its adoption. This paper presents a framework that
considers the existing learnersourcing literature, the latest insights from the learning sciences and advances in AI to offer promising future directions for developing learnersourcing systems. The framework is designed around important questions and human-AI partnerships relating to four key aspects: (1) creating novel content, (2) evaluating the quality of the created content, (3) utilising learnersourced contributions of students and (4) enabling instructors to support students in the learnersourcing process. We then present two comprehensive case studies that illustrate the application of the proposed framework in relation to two existing popular learnersourcing systems. 
\end{abstract}



\begin{keywords}
learnersourcing \sep
crowdsourcing in education \sep
student generating content \sep
human-AI partnership
\end{keywords}

\maketitle
\section{Introduction}
%
%
%
%


Our increasingly connected world is empowering learners and enabling exciting new pedagogies.  In particular, educational tools that facilitate collaboration between students can help to foster a wide range of social and domain-specific skills \citep{jeong2019ten}. The literature on computer supported collaborative learning documents a diverse range of pedagogies that have been applied for decades in many subject domains and educational levels \citep{lehtinen1999computer, roberts2005computer, kaliisa2022social}. One recent approach, derived from foundational work on contributing student pedagogies \citep{collis2002contributing, hamer2012contributing}, involves students creating and sharing learning resources with one another. Such activities have gained popularity in recent years and are associated with two broad types of benefits. Firstly, creating learning content is a cognitively demanding task that requires students to engage deeply with course concepts and exhibit behaviours at the highest level of Bloom's taxonomy of educational objectives \citep{hilton2022scalable}. Secondly, leveraging the creative power of many students can result in the rapid and cost-effective creation of large repositories of learning resources that can, in turn, be used for practice and to support personalized learning experiences \citep{singh2021whats}. 


Learnersourcing is a commonly used term to describe the practice of having students work collaboratively to generate shared learning resources \citep{kim2015learnersourcing}.  It is related to the more general task of crowdsourcing, in which tasks are outsourced to a pool of participants, often drawn from large and undefined populations, each of whom makes a small contribution to some product. The free online encyclopedia, Wikipedia\footnote{https://en.wikipedia.org}, is perhaps the canonical example of a crowdsourcing project where the number of users of the resource vastly outweighs the number contributors \citep{antin2011my}.
Crowdsourcing participants are also rarely end users in the context of tools such as Amazon's Mechanical Turk, which is a platform that harnesses human computation in the form of microtasks to solve larger problems \citep{doroudi2018crowdsourcing}.  In such cases, crowdworkers are typically paid a small fee for their contributions, although these kinds of models have drawn criticism around their exploitative nature \citep{schmidt2013good}.  In contrast, learnersourcing adopts a more human-centered focus and involves a cohort of learners studying a common subject.  The process of developing material in a learnersourcing activity is pedagogically beneficial to the learner, whereas in a more traditional crowdsourcing task it is a means to an end \citep{jiang2018review}. Learners are also inherently motivated to use the materials developed by their peers, and this is a defining feature of such pedagogies \citep{hamer2008contributing}. 


As interest in learnersourcing grows, recent work has emerged seeking to inform the design of learnersourcing tools and to guide research activities. \cite{khosravi2021charting} reflect on the experience of building an adaptive learnersourcing platform and present a series of data-driven lessons for developers and researchers.  They recommend the use of accurate and explainable consensus approaches for assessing content quality, incentives for encouraging high-quality contributions and open learner modeling to make progress visible to learners.  They also argue strongly for the need to harness the potential of artificial intelligence (AI) for improving feedback and generating effective recommendations for learners.  More recently, \cite{singh2022learnersourcing} propose a theoretical framework for studying and designing learnersourcing systems.  Their framework is centered around a set of four design questions previously used to classify crowdsourcing systems, which they augment with two questions that focus on the prerequisite skills and the learning outcomes of those who contribute new learning artefacts. They apply their framework to classify prior learnersourcing literature, and offer it as a guide for practitioners when designing novel systems.  They suggest that the two primary questions in their framework, ``what is being done?'' and ``what are contributors learning from the task?'', should be answered as the first step in any new initiatives.  When considering future directions, they highlight the complementary relationship between learnersourcing and AI.  For example, when meaningful learning activities naturally produce data that can be used for training models, which can then be used to evaluate resource quality and make personalised recommendations.


In the current paper, we propose a novel framework that captures what we see as the four essential components of learnersourcing models.  Our framework, which is organised around fundamental \emph{activities}, is complementary to that proposed by \cite{singh2022learnersourcing} which uses a learner-centric rather than an activity-centric lens.  They adapt the `what', `who', `why', and `how' design questions proposed by 
\cite{geiger2011crowdsourcing}, providing an excellent framework for classifying prior work.  In contrast, our framework considers the creation, evaluation, utilisation and instructor oversight of resources to be the four defining activities in the learnersourcing model. { In our view, one of the benefits of this activity-centric model is that each component of the framework represents a core activity in which human effort is directly enhanced through partnership with AI.  Given the rapid and transformative emergence of generative AI powered by large language models \citep{kasneci2023chatgpt, denny2023computing}, human-AI partnerships will play an essential role in the near future and thus a suitable framework to capture and describe these partnerships is needed.  Prior learner-centric frameworks that focus on issues of learner mindset and motivation are very valuable, for example to categorise literature and design incentive systems, but are less suitable for capturing the increasingly important role of generative AI \citep{chao2017rewards}.  In presenting our framework, we highlight several questions of interest for each component, and we discuss related work, current challenges and promising future directions.}

We present the framework in Section~\ref{sec:framework}, organising key questions around the `Create', `Evaluate', `Utilise' and `Oversight' components.   Section~\ref{sec:casestudies} then presents two case studies that illustrate the application of the framework in the context of two existing learnersourcing systems; namely, \toola \citep{denny2008peerwise} and \toolb \citep{khosravi2019ripple}. These case studies are complementary as \toola is considered a pioneering learnersourcing system, originally developed in 2008. In contrast, \toolb was originally released in 2019 and is still under active development, which has enabled it to benefit from state-of-the-art insights from the fields of crowdsourcing, AI and human-computer interaction (HCI). Finally, Section~\ref{sec:conclusion} provides concluding remarks about the future of learnersourcing. In particular it highlights the growing impact of AI including the use of generative language models for creating new content, deep learning models for evaluating content quality, recommendation algorithms for helping students better utilise content, and the use of experts in the loop across the four dimensions of our framework. 


\begin {figure*}[h]
\centering
\includegraphics[width=0.85\textwidth]{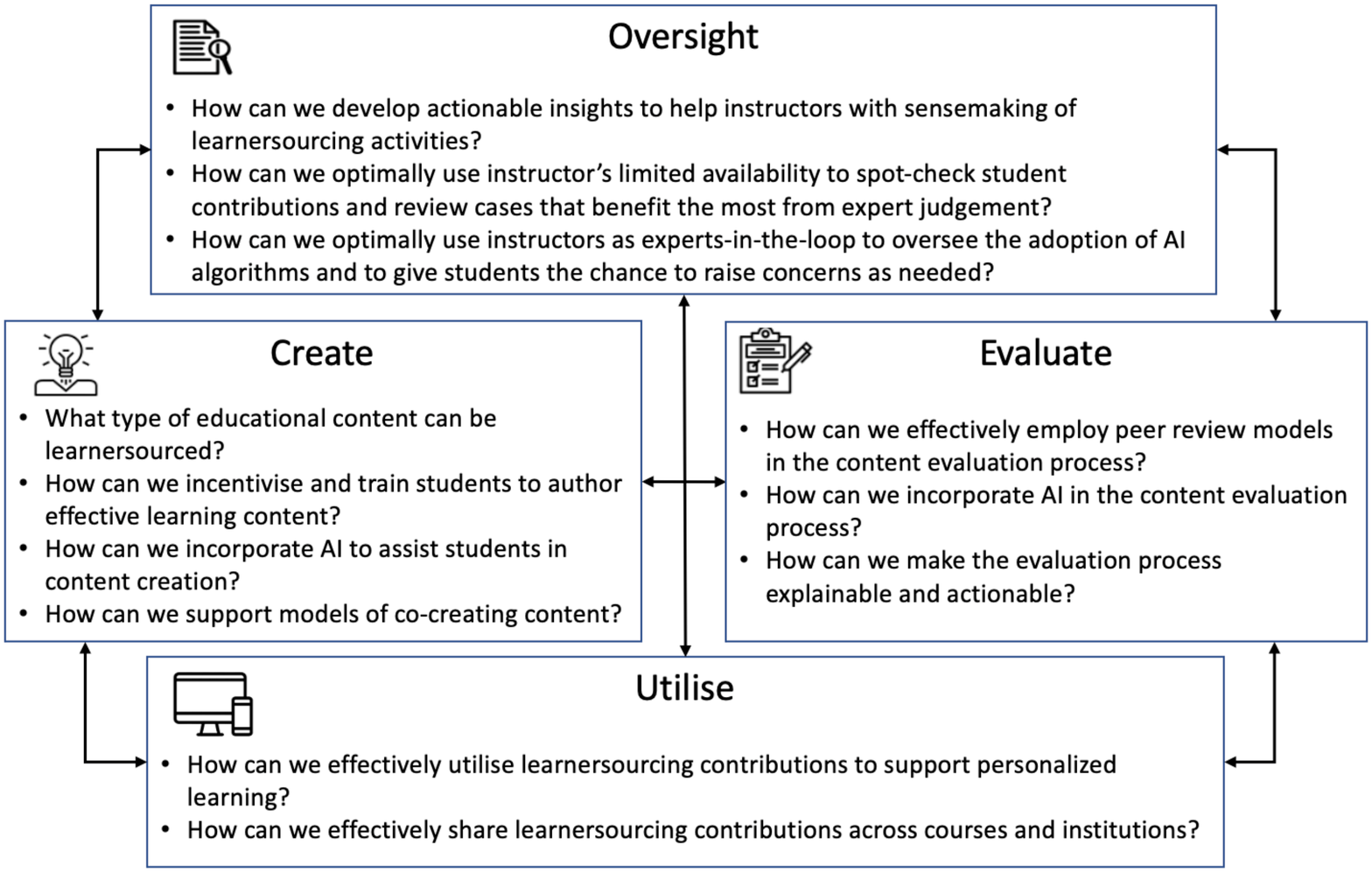}
\caption{ A learnersourcing framework related to creating novel content, evaluating the quality of the
created content, utilising learnersourced contributions of students and enabling instructors
to support students in the learnersourcing process.
\label{fig:framework}} 	
\end {figure*}

\section{Developing Learnersourcing Systems}\label{sec:framework}
This section presents the framework  that considers related work, current challenges and promising future directions around the four key activities supported by learnersourcing systems.  The framework is depicted in Figure~\ref{fig:framework}, and we organise the remainder of this section around these four components.   Section~\ref{sec:contentcreation} focuses on the authorship and creation of novel content by learners. Section~\ref{sec:contentevaluation} focuses on how the quality of learnersourced contributions can be effectively evaluated. Section~\ref{sec:contentutilisation} focuses on how repositories of content created via learnersourcing can be effectively utilised to support student learning. Finally, Section~\ref{sec:oversight} focuses on the role of educators and how they can oversee and provide support to students engaged in learnersourcing. 

{
As depicted in the framework in Figure~\ref{fig:framework}, each component interacts with the other components in various ways.  When content is created by learners, that content can be evaluated by other learners, utilised for study and review purposes, and overseen by instructors with expertise in the subject area.  In all cases, the feedback that is generated -- either explicitly through evaluation or instructor oversight, or implicitly through utilisation -- can be used by the author of the content to edit or improve their existing artefacts, or to create more effective content in the future.  Insights generated through both the utilisation and evaluation of resources by learners can feed directly into the process of instructor oversight, for example to facilitate efficient use of the instructor's time to make high impact decisions.  Similarly, aspects of expert judgement can be used to provide feedback to students when evaluating content and to utilise resources that are approved by the instructor.
Finally, learner evaluation of content can feed directly into recommendation algorithms that improve the utilisation of that content.  Conversely, utilisation patterns -- for example, aggregated data on selected options or popular answers -- can be useful data for learners when evaluating learnersourced content.

One of the key contributions 
of this paper is that it leverages recent developments in the fields of human-centred AI and AI in education to highlight opportunities for human-AI partnership across the four dimensions of the framework. In particular, it leverages recent developments from the field of AI in education such as the use of generative language models for creating educational content, deep learning models for evaluating educational content quality, learner modelling and recommendation algorithms for helping students better utilise content, and relying on instructors as experts-in-the-loop in the context of learnersourcing.  We discuss these in more depth below:

\begin{itemize}
\item The creation of educational content is a vital aspect of learnersourcing. With respect to `Create', our framework focuses on the potential range of content that can be produced, the methods of incentivizing its creation, and the collaboration of learners and AI to jointly create resources.  For example, generative AI models have proven adept at producing certain kinds of learning resources \citep{leinonen2023comparing}, and these models can be directly integrated into learnersourcing systems to improve the efficiency and the quality of the content that learners ultimately produce.  We provide a concrete example of such integration, that of generating distractors for multiple-choice questions, as part of one of our case studies in Section \ref{sec:CreateQuestionInPeerwise}.
\item Evaluating the quality of learner-generated content is an essential aspect of learnersourcing systems. Although prior work has shown that much of the content generated by students is sufficiently high quality to benefit learning \citep{kelley2019generation,walsh2018formative}, the presence of low quality content negatively affects learning efficiency \citep{moore2021human}. With respect to `Evaluate', we consider the use of peer review, AI assistance for improving the reliablity of a peer review process and the provision of explainable and actionable feedback to content authors.  
\item As students utilise content, fine-grained interaction data can be collected in real-time and provided as input to learner models that produce personalized recommendations of content, improving learning efficiency \citep{papanikolaou2014constructing}.  With respect to `Utilise', we consider the application of such learner models as adaptive engines that recommend resources, as well as the use of learnersourced content outside of institutional silos.
\item Manual review of learner-generated content by experts does not scale, and so support for efficient moderation is critical \citep{moore2023assessing}. With respect to `Oversight', we consider the provision of AI-assisted actionable insights to instructors, and the efficient use of their time to balance the cost and reliability of expert judgements.  
\end{itemize}
}

\subsection{Creating learnersourced Content} \label{sec:contentcreation}
The creation of educationally-relevant materials by students is central to learnersourcing activities, and the term \emph{student-generated content} (SGC) is often used to describe the resources produced in the context of learnersourcing \citep{snowball2017student,wheeler2008good,hardy2014student}.  Creating content can be a challenging task for students, but one which offers several distinct learning benefits \citep{singh2021whats}.  Consider, for example, some of the typical processes that a student might follow when creating content as part of a learnersourcing activity. 
To prepare for creating a specific learning artefact, a student would benefit from studying the requisite material in order to understand the concepts being targeted.  While constructing the artefact, the student may generate worked solutions or explanations of the relevant ideas, either for explicit inclusion as part of the material to be published or for their own benefit.  Either way, such explanations are beneficial and research into the self-explanation effect suggests that students who explain examples to themselves, whether prompted or not, learn more effectively ~\citep{vanlehn1992model,bisra2018inducing}. 

Students also benefit from the generative aspects of content creation.  When compared to reading content produced by others, which tends to be a passive activity, generating content leads to more robust recall \citep{crutcher1989cognitive}.  This phenomenon is related to the widely studied `generation effect' which suggests that people remember information better when they take an active role in its production \citep{slamecka1978generation}.  Originally established within the context of simple memorisation tasks, the generation effect has been shown to generalise to more complex learning materials \citep{RittleJohnson_2008_GAB,kinjo2000does,kelley2019generation} and across a variety of domains \citep{dewinstanley2004processing,scapin1982generation}. More broadly, the content creation aspect of learnersourcing is an inherently active task, and more effective for building knowledge than activities that are more passive such as listening to lectures delivered by an expert.  This active construction of knowledge is a central tenet of constructivist learning theory \citep{bada2015constructivism}, which provides theoretical support for activities such as learnersourcing.  Moreover, social constructivism posits that co-constructing knowledge with others provides an important cultural context on which personal knowledge can be built \citep{rannikmae2020social}.  In a similar vein, \cite{bredow2021flip} cite constructivism and social constructivism as providing theoretical support for the academic and interpersonal benefits of flipped learning.

Despite these well established benefits, there remains some debate as to how to best incentivise students to create learning materials \citep{khan2020completing}.  For many educational activities, a common strategy to promote participation is to reward students with some form of mark or course credit.  There is a tension here, as associating a large amount of course credit with a learnersourcing task places a burden of evaluation on instructors, whereas too small a credit can lead to resentment from some students that the reward is not commensurate with the effort \citep{doyle2019assessment}.  Rather than making it compulsory,  \cite{singh2021whats} explored giving students the choice to create multiple-choice questions (MCQs) for their peers in the context of a large massive open online course (MOOC).  They found that learners created higher quality content and valued the activity more when they could choose to participate rather than being required to do so, however fewer than 10\% of learners voluntarily created content.  While this may be suitable in the context of a large MOOC, in small class settings low levels of participation may limit the usefulness of the generated resource for practice purposes.  From a user interface perspective, the use of virtual rewards such as points and badges have shown some promise for incentivising students in learnersourcing contexts, however they tend to be more effective motivators for the utilisation rather than the creation of content \citep{yeckehzaare2020qmaps}.  For example, one study found that a badge-based achievement system had a significant effect on the number of questions answered by students for practice, but not on the number of questions authored \citep{denny2013effect}.  Subsequent work went further to establish a causal link between these types of gamification mechanics and learning outcomes, mediated by practice testing beahviour \citep{denny2018empirical}.



\subsubsection{Types of content}
In general, the learnersourcing model is broad and places no boundaries on the type of content that students can produce.  Any type of content that is associated with a course and typically produced by an instructor could be learnersourced by students.  In practice, assessment and review materials such as questions and exercises form a popular category of learnersourced resources \citep{moore2022learnersourcing}.  Their popularity may be explained in two ways.  Firstly, it leaves the production of core instructional material in the hands of experts.  This not only helps to ensure that students build appropriate knowledge prior to generating materials that assess that knowledge, but it can be viewed as less controversial than having primary learning resources generated by non-experts \citep{hamer2006some}.  Secondly, certain formats of assessment and review resources are very familiar to students, meaning they have plentiful examples to draw from.  A good example of this is the widely popular multiple-choice question (MCQ) format, which is the most common type of artefact explored to date in the context of learnersourcing.  Student-generated MCQs appear in tools such as RiPPLE \citep{khosravi2019ripple}, Quizzical \citep{riggs2020positive}, UpGrade \citep{wang2019upgrade} and PeerWise \citep{denny2008peerwise}.  Other kinds of practice questions and exercises that have been explored include complex assessments on circuits and electronics \citep{mitros2015learnersourcing}, Structured Query Language (SQL) practice exercises for database courses \citep{leinonen2020crowdsourcing} and both small-scale \citep{denny2011codewrite} and large-scale programming problems \citep{pirttinen2018crowdsourcing}.  

Although student-generated questions are common in the learnersourcing literature, a wide variety of other instructional content has also been explored.  For example, \cite{gehringer2006expertiza} describe the use of their Expertiza tool for managing student-generated contributions to their course textbook, which included students making improvements to existing explanations contained in the book and creating new examples for the concepts described in each chapter.  
Other examples that illustrate the variety of SGC in the literature on learnersourcing include subgoal labels for video tutorials \citep{kim2013learnersourcing, weir2015learnersourcing}, subgoal hierarchies for programming exercises \citep{jin2022learnersourcing}, the underlying knowledge components for assessment items \citep{moore2020evaluating}, personalised hints for engineering design problems\citep{glassman2016learnersourcing}, explanations for peer instruction questions \citep{bhatnagar2020learnersourcing}, solutions to open-ended questions \citep{wang2019upgrade} and explanations for programming misconceptions \citep{guo2020learnersourcing}.
\cite{hills2015crowdsourcing} describe the use of both learnersourced blogs and videos in a psychology course, where students generate content that aligns with their personal interests.  Students were tasked with collecting existing resources from their everyday experiences and curating them on a blog, as well as producing novel content in the form of persuasive videos promoting pro-social messages.  


\subsubsection{Promoting high quality content}
In order to create high quality content within a learnersourcing task, students need both domain-specific knowledge and task-specific knowledge \citep{devine1995domain}.  The former is typically developed through engagement with course learning materials and other curricula resources.  Alongside this core disciplinary knowledge, students also need to understand how to construct high quality learning resources.  Depending on the type of artefacts being learnersourced, this may include knowledge of how to construct effective MCQs, or how to generate useful hints or helpful explanations \citep{snow2018discursive}.  This task-specific knowledge may be taught directly by the instructor or incorporated as part of the learnersourcing system \citep{doyle2019assessment}. 

Various approaches for the instruction of task-specific knowledge have been reported. \cite{doyle2019assessment} describe a learnersourcing activity involving MCQs where guides for constructing effective MCQs were made available to students on the course learning platform.  However, they observed that these guides were infrequently consulted by students, despite the fact that students complained of the need for more support on how to construct questions.  As a result, they recommend teaching the principles of good MCQ design explicitly, and providing examples of both good and weak questions to illustrate these principles.  
\cite{bates2014assessing} describe a deliberate approach to prepare students for generating MCQs, involving a 90 minute tutorial consisting of several elements.  These included a content-neutral quiz to familiarise students with the terminology of an MCQ (e.g., stem, options, distractors), a self-diagnosis quiz to guide students towards a learning-orientation rather than being performance focused, a representation of Vygotsky's ``Zone of Proximal Development'' \citep{chaiklin2003learning} to challenge students to author questions of high cognitive value, and question exemplars.  To assess the value in this deliberate approach to teaching task-specific knowledge, the authors evaluated the questions authored by students finding that 75\% passed a set of quality criteria including explanation detail, distractor plausibility, question clarity and cognitive level. 
An even more rigorous approach to scaffolding the MCQ creation process was documented by \cite{hilton2022scalable}, in which students were introduced to increasingly complex tasks in five distinct steps over a period of 10 weeks.  These began with students constructing short statements that are either true or false, to practice writing concise statements targeting a single topic, and then suggesting improvements to MCQs they found online before progressing to authoring their own MCQs.  The authors see their scaffolding of task-specific skills as essential to the broad and deep conceptual benefits they observed as a result of the learnersourcing task. 

One commonly reported challenge with the learnersourcing of assessment items such as MCQs is that many of them end up being simple recall questions \citep{moore2021examining}.  For example, in a study by Bottomley and Denny that used \toola for learnersourcing MCQs, more than half of the student-generated questions were classified at the lowest level of the revised Bloom's taxonomy \citep{krathwohl2002revision}.  In their study, students were provided exemplar questions but were not explicitly instructed on the learning objectives or the cognitive processes associated with them as described by a taxonomy like Bloom's.  The extent to which such instruction is helpful is not clear and is likely highly contextual.  In some situations, the use of exemplars alone has proven effective for training novices in other crowdsourcing contexts.  For example, 
\cite{doroudi2016toward} explored the effects of different training strategies on novices in a crowdsourcing task, finding that the provision of expert examples outperformed other training strategies.  Similarly, in the context of learnersourcing subgoal labels, \cite{choi2022algosolve} present learners with good examples of subgoal labels in a `warm-up' training phase before they are asked to create their own labels.  Also in the learnersourcing literature, \cite{huang2021selecting} reported good success explicitly teaching students about Bloom's taxonomy and its application to assessment items. Students practiced assessing questions according to Bloom's taxonomy before creating their own, and the authors found that a selection of the student-generated questions performed as well as questions generated by academics on summative exams.

Moving beyond task-specific knowledge,  \cite{lahza2022effects} explore the benefits in a learnersourcing context of scaffolding self-regulated learning behaviours. In a controlled study, they investigated the benefits of explicitly prompting students to plan their work before creation, self-monitor during creation, and self-assess after creation.  Although these metacognitive scaffolds have robust theoretical benefits, in practice the authors found that they increased task complexity and completion time, without any significant improvement in the quality of the content students produced.  In general, the extent to which different types of training resources are effective, and the tradeoffs they present in terms of time and scalability of instruction, is currently under-explored in the learnersourcing literature.






\subsubsection{AI assistance in content creation}
Creating high-quality and novel content is demanding, and not all students engage well with the generative aspects of learnersourcing \citep{moore2021examining}.  Indeed, prior research has shown that students are often more inclined to use and evaluate resources that are created by others rather than expend the effort needed to create high quality content of their own \citep{singh2021whats, pirttinen2022can}.  Using machine learning techniques to automatically generate novel content, at least in a draft form that a student could refine, is therefore a promising approach for scaffolding the learnersourcing process with AI.

Large language models (LLMs) have recently emerged and proven very effective at generating realistic human-like content.  Models like GPT-4 \citep{openai2023gpt4} and Open AI's Codex \citep{chen2021evaluating}, which are respectively fine-tuned to produce natural language text and source code, have received a great deal of attention.  Such models are not limited to textual output, with models such as DALL$\cdot$E \citep{ramesh2022hierarchical} being able to produce extremely creative artistic images from natural language prompts.  These models tend to be very good at ``few shot'' learning, in which the input prompt includes one or more contextual examples which leads to the generation of a novel output.  Early evidence suggests that these models are very good at creating educational content \citep{wang2022towards}.  Recent work by \cite{drori2021machine}
applied both GPT-3 and Codex to generate novel university-level mathematics problems with explanations, and to solve them with equivalent success rates to humans.

In natural language processing and machine learning, it is common to use questions to train models to both generate higher quality questions and enable them to answer them with higher accuracy \citep{wang2021towards}. Naturally, learnersourcing could provide a great source of training data for these applications. Such models may be useful in assisting with the content creation phase of learnersourcing.  For example, a learner could provide an initial prompt to the model, which may include examples and other contextual priming information, and would then be able to evaluate and refine the content produced by the model.  Such an approach places a greater emphasis on content evaluation than on content creation, and may improve the efficiency with which large-scale resources can be produced. \citep{sarsa2022automatic} use the term ``robosourcing'' to describe this augmentation to the traditional learnersourcing model. 
In general, we expect that as generative AI models improve and become more deeply embedded in educational contexts, there will be a shift in emphasis with respect to the application of higher-order thinking skills.  Creating novel content, traditionally seen as an important high-level skill, will become relatively less important when compared to evaluating and critically analysing existing content, given the ease with which AI models can produce new content.  Similarly, the importance of developing certain lower-order, foundational skills may become relatively less important given the availability of AI models that provide suitable support \citep{denny2022robosourcing}.

\subsubsection{Co-creation models}
Most existing learnersourcing systems expect individual learners to generate and contribute complete learning resources (for example, in the case of MCQs, this would typically include the stem, a set of distractors, and an explanation or sample solution) \citep{singh2022learnersourcing}.  This requires a large investment of time from a learner and can be error prone as individual parts of the artefact are not reviewed by others before being assembled.  In contrast, many crowdsourcing systems tend to be structured around small micro-tasks (e.g., labelling one image \citep{chang2017revolt}) which permit useful contributions with little time and effort.  Moreover, these systems often support a co-creation model where individuals can collaborate on the same artefact.  Examples of this type of model include editors in Wikipedia being able to directly modify content produced by others, and forums like StackOverflow allowing users with sufficient reputation to edit questions and responses generated by other users.  Another early example of this idea was the novel wordprocessor plugin by \cite{bernstein2010soylent}, called Soylent, which permitted crowdworkers to make small proof-reading edits to a document, including minor formatting and wording changes.

Learnersourcing can make use of co-creation models to empower individual learners to contribute in a variety of different ways.  \cite{singh2021whats} explored the factors that influenced learners' decisions to create content when it was optional to do so.  Lack of time, low confidence and lack of interest were the primary reasons cited for choosing to not create content.  As a result, they advocate for co-creation models that allow for tiered contributions such that students with little time or less confidence could contribute in more granular ways.  In the context of their study, which involved authoring MCQs, they suggested that rather creating a complete MCQ, learners could provide a set of distractors or write an explanation.  Recent work by \cite{kim2022dynamic} explored this very idea through a modularized approach to learnersourcing MCQs.  In their model, individual components of a question, such as a single option or a stem, could be authored and refined by learners thus providing flexibility for different learners to contribute in ways that suit their interests.


 
 
 

\subsection{Evaluating learnersourced content}
\label{sec:contentevaluation}
One of the effects of learnersourcing is that it makes it relatively easy to develop large repositories of SGC. While strong evidence from previous work suggests that a large portion of the SGC is of a high-quality and meets rigorous judgmental and statistical criteria \citep{walsh2018formative,galloway2015doing}, it also suggests that students commonly create resources that are ineffective, inappropriate, or incorrect  \citep{tackett2018crowdsourcing,denny2009quality,bates2014assessing}. As a consequence, to effectively use SGC repositories,
there is a need for separating  high-quality from low-quality resources. One approach is to engage instructors as experts in evaluating the quality of the resources; however, the instructor-led quality evaluation is not scalable and can be expensive due to the potentially large size of the repositories (Section~\ref{sec:oversight} explores plausible methods of optimally using instructor's limited availability towards evaluating content). This section explores two alternative approaches of employing human or machine computation for evaluating the quality of the resources. Primarily, we first explore the possibility of incorporating co-regulation models of a peer review process where students are engaged in  assessing the quality of resources authored by their peers. We then explore how AI methods can be incorporated to help with the assessment of the quality of the resources. We finally discuss the need for accurate, fair and transparent evaluation methods, regardless of whether they are done by humans or machines. 

\subsubsection{Peer review models for evaluating quality}
Peer review is a well established model for evaluating quality, often employed for academic publishing \citep{tennant2018state}. 
To determine the suitability of the peer review process in the context of learnersourcing, it is important to first consider whether engaging students in evaluating content is beneficial to their learning.  If such benefits are absent,  ethical issues arise with respect to utilising students as cheap labour to reduce the workload of instructors \citep{zdravkova2020ethical}. There is however a general consensus that engaging students in peer review has many benefits \citep{Nicol2014}. These include enabling them to improve their comprehension of the content \citep{li2010assessor}, develop evaluative judgement \citep{tai2018developing,gyamfi2022effects,Khosravi2022EJ} and a sense of accountability \citep{kao2013enhancing}, and improve their writing \citep{polisda2017peer} and ability to provide constructive feedback \citep{lundstrom2009give}. 

An equally important question to consider is whether students have the capacity to effectively evaluate the quality of peer-created resources. Prior work suggests that students by and large have the ability to accurately evaluate the quality of learning content  \citep{galloway2015doing,tackett2018crowdsourcing}. Having students as evaluators also addresses expert blind spot challenges as they would evaluate the effectiveness of a resource based on their own previous misconceptions \citep{nathan2001expert}. However, as experts-in-training, their judgements cannot wholly be trusted \citep{abdi2021evaluating}. A common solution, which is also incorporated in academic publishing, is to rely on the wisdom of the crowd rather than one person by employing a redundancy-based strategy and assigning the same reviewing task to multiple users \citep{reily2009two}. Other strategies such as utilisation of rubrics \citep{gyamfi2022effects, Gyamfi2022}, exemplars \citep{carless2018developing}, guides on providing effective feedback \citep{darvishi2022incorporating} and comparative judgement where students choose the `better' of two pieces of work \citep{cambre2018juxtapeer, palisse2021comparative}
have also been shown to be an effective method for helping students develop evaluative judgement.

The peer evaluation process can determine the way in which resources are shared.  For example, one approach is to allow optional and subjective ratings on artefacts that have already been made available to all students \citep{denny2008peerwise}. In such cases, aggregate ratings can be used to help students search for artefacts that have received higher quality scores \citep{denny2009quality} or be used to support personalisation mechanics \citep{williams2016axis}. A more restrictive approach might follow the academic publishing process in which artefacts are judged by a subset of users using  multi-criteria rubrics. In this case,  reviews are used to determine whether a resource is of high enough quality to be approved and shared with other students or if it lacks the required quality and is to be rejected and sent back with feedback to the author \citep{khosravi2019ripple}. A workflow was developed by \cite{lee2020question} and deployed into a system known as Questionable, that allowed students to author and review questions that were then presented to their peers. Using this workflow, students would leave feedback regarding a question generated by their peers, indicating how it might be improved or assessing the quality and usefulness of it. These reviews were then presented to course staff, that used them to quickly determine the quality of the question or make any necessary changes to them. Ultimately this allowed the course to make quick use of the questions and ensure only the highest quality ones were being shown to the students.

\subsubsection{AI in content evaluation}
While the notion of following an academic publishing model for evaluating SGC is time-effective and supports student learning, it does introduce new challenges where AI can be of assistance. Here we provide two examples; firstly, unlike the publishing model where a meta reviewer makes the final call, in the case of SGC, it is impractical to expect the instructor to meta-review possibly thousands of artefacts that are being created in their course. Therefore, the process of deciding whether an artefact is to be approved or rejected based on multiple reviews needs to be automated. This  raises a new problem commonly referred to as the consensus problem \citep{10.1145/3597201}. In the absence of ground truth, how can we optimally integrate the decisions made by multiple people towards an accurate final decision? Traditional consensus approaches rely on general statistical aggregations, such as the arithmetic mean, median, or majority vote \citep{zheng2017truth}. However, previous studies have shown that there is an evidential difference in the judgemental ability of students \citep{abdi2021evaluating}. Inspired by work from crowdsourcing \citep{zheng2017truth}, an interesting approach has been to use machine learning models to infer the reliability of a reviewer or a review such that the consensus model can put more weight on decisions made by reliable reviewers \citep{darvishi2021employing}. Another challenge with relying on student reviews and feedback relates to students' failure in providing high-quality feedback, which leads to substantial negative consequences such as lowering standards \citep{yeager2014breaking}, reducing trust in the outcome \citep{carless2009trust}, and making reviewees less likely to revise their work \citep{sommers1982responding}. Here, inspired by advances in natural language processing to evaluate the quality of a review \citep{negi2016study,devlin2018bert}, a possibility is to develop quality control functions that automatically assess the quality of the submitted feedback and ask students to improve, if necessary \citep{darvishi2022incorporating}.

There have also been various attempts to automatically evaluate the quality of MCQs and more broadly educational artefacts. Metrics such as the discrimination index from classical test theory have been used for decades for identifying the quality of MCQs  \citep{malau2014using}. However, a limitation of this method is that it requires large quantities of data on student responses to items. Therefore, it cannot be used to evaluate the quality of new questions. {The 2020 education challenge \citep{wang2021results} from the Conference on Neural Information Processing Systems (NeurIPS) has started a new wave of using advanced AI models for the automatic determination of the quality of MCQs (Task 3 of the challenge). Three teams were announced as co-winners of this task, each achieving an 80\% agreement with human evaluators’ judgements. The approaches by two of the three winning teams Shinahara \& Takehara and TAL Education presented solutions that computed explicitly-defined features based on the hypothesis that high-quality MCQs are appropriately difficult, readable, and have a balance among answer choices \citep{wang2021results}. Interestingly, the other winning approach by McBroom \& Paassen did not use any complex feature engineering and had the very simple hypothesis that the quality of an MCQ correlates with the confidence of students answering it \citep{McBroom2020NeurIPS}}. They argued that high student confidence implies that the question is clear and unambiguous. In addition, they argued that the Dunning-Kruger effect \citep{dunning2011dunning} may result in students holding key misconceptions by reporting high confidence in incorrect answers if the question clearly addresses this misconception. More recently, \cite{ni2021deepqr} propose DeepQR that, alongside computing explicitly-defined features, uses a 2-layer transformer encoder to consider semantic features, which are designed to capture relations between different question components. Compared to six existing models, DeepQR was able to more accurately identify questions that were low or high quality. Another study trained a state-of-the-art language model, GPT-3, on learnersourced questions to classify the quality of a question as low or high and the cognitive level, according to Bloom’s revised taxonomy \citep{krathwohl2002revision,moore2022assessing}. They then had the model classify student-generated short answer questions to automatically classify their quality and cognitive level.

It is worth highlighting that the majority of existing work has focused on automatic evaluation of MCQs \citep{kurdi2020systematic}. However, the advancements in natural language processing (NLP) on pre-trained language models such as Bidirectional Encoder Representations from Transformers (BERT) \citep{devlin2018bert} and its extended models make it possible to generate and automatically detect the quality of content, which can also be tailored toward educational artefacts. Another limitation of the current AI-based evaluation methods is that they focus mostly on explicitly-defined and semantic features rather than the correctness of the content. An interesting future direction is studying how AI and students can collaborate on content evaluation where the correctness of the content is examined by students and the readability and flow is examined by AI. 

\subsubsection{Explainable and actionable evaluation methods}
Much of the existing work on evaluating learnersourced content has focused on just separating out high and low quality resources \citep{abdi2021evaluating}. However, an important aspect of engaging students in learnersourcing and evaluating their work is to help students not only improve their ability in content creation but also to help them develop their ability to monitor, evaluate and regulate their learning so that they can revise and enhance their created content. This is of particular importance in the cases where learnersourcing activities are tied to assessment in which the assessed quality of a resource impacts student grades \citep{singh2021whats}. In the case of using peer review, an interesting future direction is to study how best practices from feedback literacy in terms of engaging authors and evaluators in dialogue \citep{ajjawi2017researching} and feedback loops \citep{carless2019feedback} can be applied in learnersourcing. In the case of AI-based evaluation, the current use of deep learning methods for assessing the quality of educational artefacts such as MCQs has shown to be accurate and close to human judgement; however, the models operate as black boxes, providing no justification for their decisions.  An interesting future direction is to study how explainable AI methods in the context of education \citep{khosravi2022explainable} can be applied to learnersourcing. 



\subsection{Utilising learnersourced content}\label{sec:contentutilisation}
A side benefit of engaging students in content creation and evaluation is that it enables the development of large repositories of learning resources, which can be shared with students to provide practice opportunities \citep{singh2022learnersourcing}. For instance, a plethora of studies describe the generation of large question banks across a variety of domains, and the production of millions of MCQs, that have been subsequently used for practice purposes \cite{moore2022learnersourcing}. However, the utility of the developed repositories has often been limited to the courses where the content was originally produced, which restricts the life and impact of the created content. In addition, typically students only have access to simple search and filtering functionality for the selection of practice questions. This may lead to students spending their time ineffectively on resources that are targeted toward an average student of the class rather than focusing on their knowledge gaps \citep{koedinger2013new}. This section explores how these limitations can be addressed so that SGC repositories can be more effectively utilised. We first discuss approaches for supporting personalisation in engaging with SGC repositories. Then we discuss how SGC repositories can have an impact beyond the course of their origin and to be utilised in future offerings of different courses across institutions.

\subsubsection{Personalisation in engaging with content repositories}
Learnersourcing yields student-generated artefacts, such as questions, but the process of students interacting with such learnersourcing activities and systems also generates auxiliary user-item-outcome data that can be leveraged for developing learner models \citep{abdi2022learner}.  A learner model is an abstract representation of a student's knowledge state. They are a core component of adaptive educational systems that provide students with customized learning paths and adaptive feedback based on their learning process \citep{koedinger2013new}. In the case of modelling learners in learnersourcing, \cite{moore2022leveraging} and \cite{abdi2019multivariate} have utilised data collected from students' interactions with a repository of SGC to model the knowledge and skills required to solve problems in the context of chemistry, programming and relational databases courses. As the process of creating and evaluating content also leads to student learning,  data captured during the learnersourcing activities can also be incorporated into these learner models akin to how student data from problem attempts is currently utilised. Empirical results from two studies that present learner models, which utilise data from students’ learnersourcing tasks, demonstrate that these models outperform learner models that only utilise traditional assessment data \citep{abdi2020modelling, khosravi2021charting}. Educational recommender systems can make use of a learner model to recommend learning content, which can optimize student learning and their time spent. For example, \cite{khosravi2017} present a new recommender system that recommends resources from an SGC repository. They found their approach to be able to adequately provide personalised recommendations for students who have previously used the platform as well as cold start users who are new. \cite{abdi2020complementing} report that complementing recommendations on content from an SGC repository with a learner model lead to an increase in student engagement and a positive effect on students' perceptions of the quality of the recommendations.  

Another way to support personalisation, which is commonly referred to as step-loop or inner-loop adaptivity, is to enable an adaptive instructional system to provide support to learners within a particular learning task (e.g., hints or explanations they receive) based on their performance \citep{aleven2016example}. A project by \cite{glassman2016learnersourcing} designed a learnersourcing system to provide hints to students working through a college-level programming course. As students worked on the problem, they were able to automatically receive student-generated hints that would continually update or they could elect to only receive the hints when they requested them in a just-in-time fashion. This personalization allowed students to leverage hints in a manner that fit their preferences. They proposed two models for their hints. The first is a push model, where the student-generated hints are presented to learners and constantly updated. The second is the pull model, where learners only receive hints when they request them. These two model approaches were extended by \cite{singh2022learnersourcing} and applied to broader learnersourcing applications. They indicate that the push model can be leveraged when the learnersourced artefact is intended to help the learners in the problem-solving process. On the other hand, the pull model should be utilized when the learners have completed a problem, but they might be seeking a more optimal solution. Another related system known as SolveDeep leverages student-generated sub-goals on algebra problems to provide feedback to other learners’ solutions, akin to providing a hint \citep{jin2019solvedeep}. They found that participants effectively leveraged the subgoals generated by other learners, which helped them effectively and efficiently solve several algebra problems. Another example of supporting inner-loop adaptivity in learnersourcing comes from the popular Axis system \citep{williams2016axis}, which enables students to generate explanations for math problems and then used adaptive multi-armed bandit algorithms to deploy the optimal explanations to students. They found that explanations delivered to students led to higher learning gains than a majority of the existing explanations previously used for the problems.

\subsubsection{Sharing learnersourcing contributions}
A present challenge in the learnersourcing space is the sharing of SGC across courses, institutions, and ultimately to a broader audience. While efforts in the open educational resources (OER) space provide insights into the dissemination of instructional and assessment content, it is often content that is created by professional instructors that intend to share their content \citep{wiley2014open}. Potential issues around copyright and the leaking of question bank answers create challenges on how we can readily share these materials in a way where they can effectively be used for both formative and summative assessments of student learning. Platforms such as OpenStax and ASSISSTments, popular OER platforms, attempt to address such challenges by requiring instructors to verify their identity before accessing the materials \citep{pitt2015mainstreaming, heffernan2014assistments}.
Researchers and practitioners continue to expand their efforts in sharing learnersourcing contributions from their courses and systems. For example, \cite{quintana2018mentor} have students in a data science course develop questions that were then utilised by students in future semesters of the same course. In the context of MOOCs, \citep{kim2014data} and \citep{weir2015learnersourcing} learnersourced labels for instructional videos that were leveraged by learners across multiple courses. \citep{kay2020student} explore how students sharing resources and learnersourcing across multiple institutions can effectively be handled. Nevertheless, approaches and adoption of methods that support sharing learnersourced contributions are under-researched and -explored.

Even when the learnersourced content is created at a single institution, factors such as the student demographics and their location may impact how SGC is utilised \citep{moore2023who} \cite{morales2020nationality} investigated how students at the same institution, but split geographically between campuses in the United Kingdom and China, perceived the SGC of their peers. They found that when the students were identifiable, it had a significant impact on how their content was accessed and rated by their peers, such as students intentionally avoiding content created by classmates of certain nationalities. Other work in the space had more positive results, as \cite{denny2012case} had students in an introductory programming course at an institution in New Zealand generate learning resources for students in a similar course in Canada. The results indicated that this cross-institutional learnersourcing worked as well as it previously did within-institution and students from both institutions indicating that they prefer their contributions be shared more widely.

\subsection{Overseeing the creation of learnersourced content} \label{sec:oversight}
The role of the instructor in the learnersourcing process can vary depending on what system they might be using, how they elect to utilise the SGC in their course, or what they want to gain by having students participate in learnersourcing \citep{khosravi2021charting}. No matter their role, they have a form of oversight in the learnersourcing process that enables them to gain insights and ultimately facilitate student learning.  While in theory learnersourcing systems can operate without the presence of an instructor, academic oversight of the creation and evaluation process can serves as a demonstration of reliability, providing assurance to both educators and students that the system is trustworthy and dependable and to encourage high-quality contributions and peer reviews \citep{darvishi2022assessing}. It also provides useful insights into the student learning process that instructors can act on, such as modifying their curriculum based on where students have difficulty or having students create questions over a particular content area. Additionally, as the size of both in-person and online courses increases, resulting in increased student-generated contributions, it can be challenging for instructors to effectively use their time in evaluating and utilising these resources \citep{ji2022qascore}. While the role of the instructor is under-explored in the learnersourcing literature, work in the related fields of learning analytics, crowdsourcing, and machine learning can provide valuable insights that can be adopted in this context.

    \subsubsection{Learning Analytics and Actionable Insights}
Imagine a tool that is supporting engagement with learnersourcing for a class with over 500 students. How can the teaching team make sense of students' engagement and performance and how can they effectively facilitate learning? Data, in addition to the content, generated from learnersourcing activities can readily be leveraged by analytic systems and learning analytics dashboards \citep{matcha2019systematic} with actionable insights \citep{jorno2018constitutes} to help instructors make sense of student learning and intervene with pedagogical interventions as necessary. 

This raises the question of what metrics and analytics can be obtained from learnersourcing that instructors might find insightful. General metrics such as the number of logins, resources created, evaluated and attempted are readily available (see the case studies from Section~\ref{sec:casestudies}) which might give instructors a sense of student engagement. Research has shown that student participation and completion are key indicators to increased learning \citep{moore2022ICLS2}. Studies also show that interaction with learnersourcing activities is not constant across student populations. The 90-9-1 rule was noted to apply in learnersourcing \citep{khosravi2021charting}, stating that 90\% of users are lurkers, 9\% create some content, but the majority is created by 1\% of the student population, which is similar to participation in forums within online courses. Engagement analytics may lead to instructor actions of congratulating high-achieving students and sending nudges and reminders to inactive students \citep{plak2023raising}. Systems, such as RiPPLE, have also provided a type of learning analytics dashboard that can be leveraged by instructors to view questions students are struggling with or content areas where students may require additional support \citep{khosravi2021charting}. Through the use of such analytics, students could then be encouraged to generate content in these troublesome areas, which could result in them thinking critically about the area, while also having the benefit of generating additional practice questions. In addition, performance data on students' interactions with assessment items can be leveraged to model student learning \citep{abdi2020modelling, kay2020student} and to infer the quality of assessment items \citep{huang2021selecting}. More fine-grained data about the sequence of activities conducted by students can be leveraged to identify underlying tactics and strategies \citep{matcha2020analytics} that are used by students while engaged in learnersourcing \citep{lahzatactics2022}. Another use case of actionable analytics for learnersouricing systems is to help instructors find resources or peer-review cases for being spot-checked which is discussed next.

\subsubsection{Evaluating/Spot-Checking}
Peer evaluation is an essential component of learnersourcing, as students commonly review the contributions of their peers within the same course or even across multiple institutions \citep{darvishi2021employing}. However, peer evaluation is often susceptible to students being unmotivated to evaluate the work of their peers in a diligent manner \citep{liu2016learning}. To make the peer review process more reliable, one potential approach is to utilize spot-checking, where an instructor or tutor evaluates some assignments and offers a reward to students who grade in a similarly diligent manner \citep{cambre2018juxtapeer}. Various metrics may be utilised for optimally determining the resources which would benefit from spot checking the most. \cite{gao2019incentivizing} show that even random spot checking can incentivize reviewers to be more diligent. \cite{wang2020optimal} take a game theoretic approach to suggest optimal spot-checks to maximise the evaluation accuracy of reviews. \cite{darvishi2022trustworthy} use a range of human-driven metrics (e.g., high-disagreement in moderation evaluations, a high ratio of downvotes in comparison to upvotes) and data-driven metrics (e.g., assessment items that have a low discrimination index or questionable distractors where the popular answer is not the one proposed by the author) in the context of learnersourcing to categorise resources into having high, medium, low or no priority for being reviewed.

It is worth noting that an alternative or complementary strategy to spot checking is to incorporate calibration submissions \citep{wang2020optimal} for which the true grade is known. These submissions can be used for training where reviewers would have access to how an expert would have graded the task. A side benefit is that they can be used to infer the reliability of a reviewer. Relatedly, previous work by \cite{hamer2005method} has also explored a technical approach to this spot-checking problem, using algorithms to calibrate peer review scores automatically. Their approach identifies ``rogue'' reviewers that appear to assign their scores arbitrarily. Scores provided by these reviewers are weighted lower when the computed aggregate score for the artefact under review is determined.

\subsubsection{Human-in-the-loop}

When it comes to AI and machine learning, human judgement is still needed to make sense of or improve the results of the AI \citep{divate2017automatic}. This is one area where learnersourcing can be leveraged and the proposed framework can provide the insights to make it work. Human judgement and human-in-the-loop becomes even more important when we move away from traditional measures of educational achievement and focus on issues around fairness, accountability, transparency and ethics (or FATE), which was the focus of a recent special issue journal that had a common theme among the papers -- that humans are still necessary ``in the AI loop'' to manage these issues \citep{woolf2022introduction}. 

In a number of systems, AI guides the use of learnersourced content with human input. One example is AXIS \citep{williams2016axis}, which uses multi-arm bandit methods to choose which student generated explanations to give students who need help. In this system, there is the opportunity for students to provide ratings on the explanations. These ratings and the future success of students on similar problems are combined to give the highest rated and best performing explanations a higher probability to be seen by future students.

One emerging area where human oversight may be needed in future learnersourcing systems is with the use of generative language models.  
\cite{sarsa2022automatic} propose the idea of robosourcing, where content generated by language models can be used as a starting point for students to accelerate the learnersourcing process.  On the one hand, the increasing automation supported by such models may suggest less need for human input, but there is a need for caution.  In their review of the opportunities and risks offered by foundation models, \cite{bommasani2021opportunities} explicitly warn against the removal of teachers from the loop. Large language models are trained on broad data produced by humans, and thus are known to suffer from biases similar to humans \citep{chan2023gpt}. Using automatically generated content as the basis for learnersourcing tasks runs the risk of perpetuating some of these biases.  We see a human-in-the-loop approach, involving both students and instructors, as essential for moderating such biases and for improving and tailoring the performance of the underlying generative models for suitability in learnersourcing contexts.






\section{Case study}\label{sec:casestudies}
In this section, we present two comprehensive case studies that illustrate the application of our proposed framework. Section~\ref{sec:peerwise} discusses a pioneering learnersourcing system, \toola, and Section~\ref{sec:RiPPLE} describes a state-of-the-art learnersourcing system, \toolb, in the context of the proposed framework.
\subsection{\toola}\label{sec:peerwise}
\begin {figure*}[h]
\centering
\includegraphics[width=0.5\textwidth]{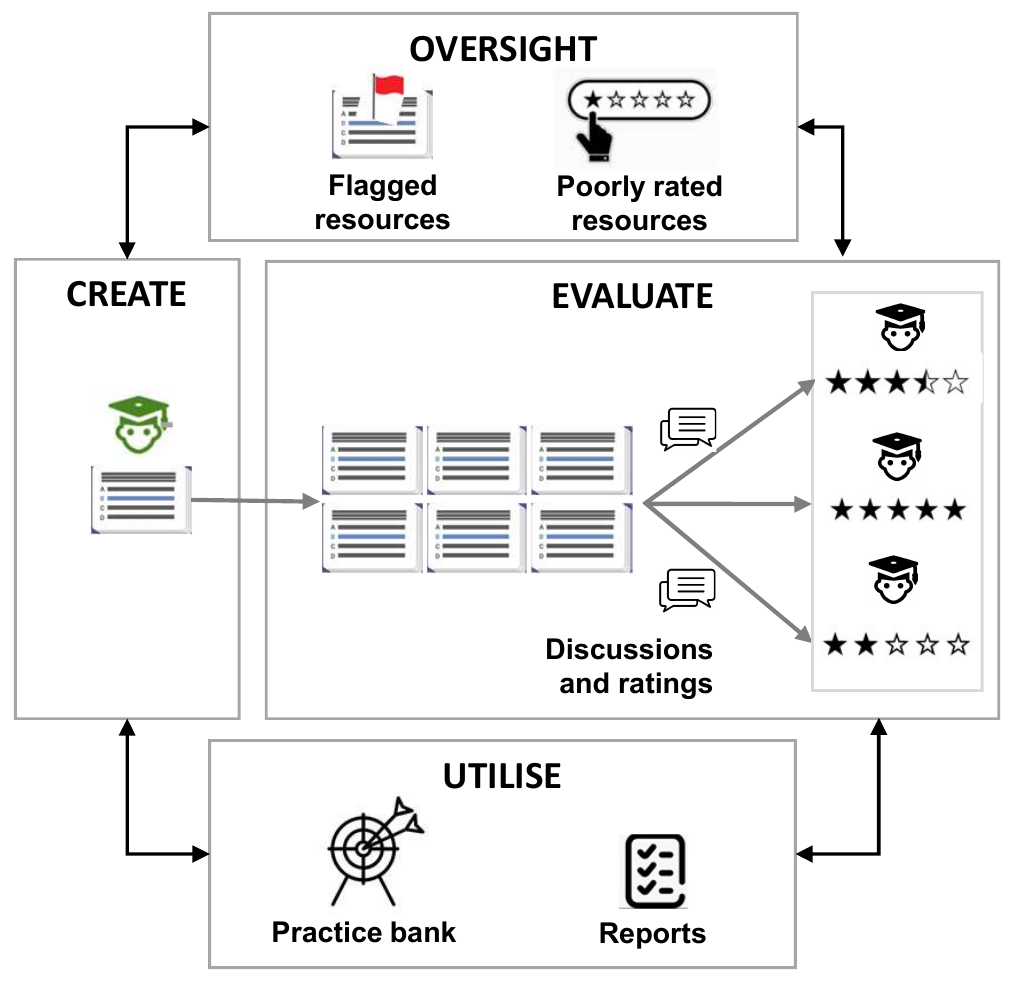}
\caption{ Overview of \toola.  \label{fig:peerwiseframework}} 	
\end {figure*}  

\toola \citep{denny2008peerwise} is a web-based learnersourcing tool, first developed in 2008, which supports students in creating, publishing, answering and discussing MCQs.  As of the time of writing, approximately 7,000,000 questions have been published by students at 3,000 institutions worldwide\footnote{peerwise.org} 
Organisationally, the content within \toola is arranged hierarchically into ``institutions'' and ``courses''.  Typically, an instructor would create a new course repository associated with their institution, and then grant their students access to that repository.  Instructors and students use the same interface, although additional features are available for instructors to provide oversight, such as running basic usage reports and managing access permissions.  Figure \ref{fig:peerwise_menu} shows an example from the perspective of a student of the main menu for one course repository.  Questions are organised with respect to whether they have been authored or answered by the student.  The user interface of \toola keeps student identities anonymous, which is a deliberate design choice that has been shown to reduce certain kinds of biases in online learning environments \citep{morales2020nationality}. 

Instructors can also access fine-grained data for their courses, which includes timestamped records of all student interactions.  The availability of this data has facilitated the work of educational researchers exploring various aspects of learnersourcing.  To date, 123 articles by 262 distinct authors have been published using data collected by \toola  \footnote{https://peerwise.cs.auckland.ac.nz/docs/publications/}. 
Much of this work has focused on learning effects.  For example, Kay, Hardy and Galloway used multilevel modelling to analyse data from 3,000 students over three years and across 18 physics, chemistry and biology courses at three UK universities \citep{kay2020student}.  When controlling for prior ability, they found a significant positive association between students' engagement with \toola and their performance on end of course exams.  They conclude that \toola offers a ``low-risk'' and ``low-cost'' intervention that supports student learning and, more generally, that learnersourcing of course material in a structured way can provide measurable educational benefits.

Figure \ref{fig:peerwiseframework} provides an overview of the operation of \toola with respect to our framework.  We elaborate on each part of the framework in the following sections. 

\begin {figure}[h]
\centering
\includegraphics[width=0.7\textwidth]{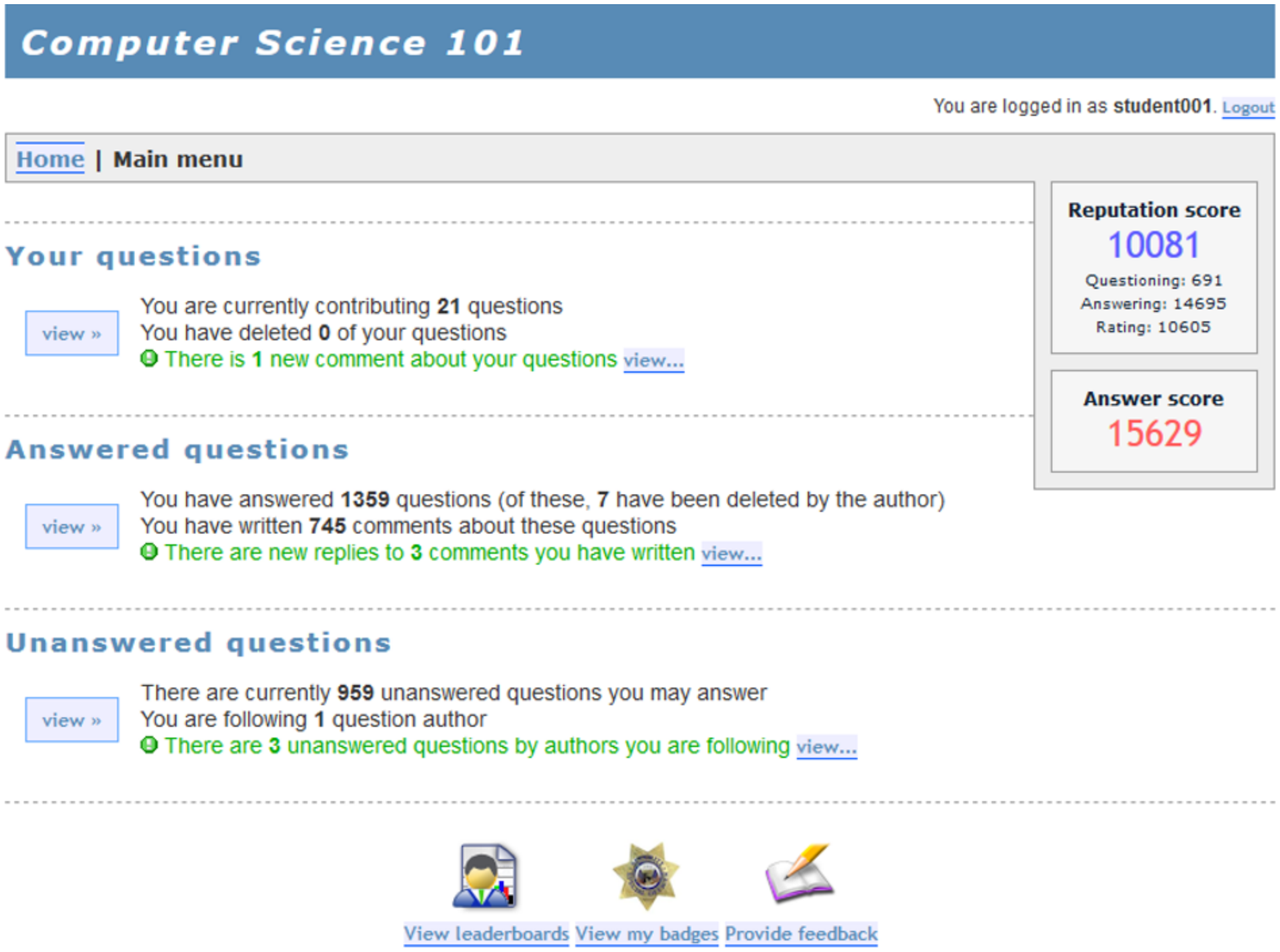}
\caption{ Main menu of \toola \label{fig:peerwise_menu}} 	
\end {figure}  

\subsubsection{Create in \toola}
\label{sec:CreateQuestionInPeerwise}
The multiple-choice question (MCQ) format is simple, widely used in practice, and is a familiar format
to most students.  When constructing an MCQ in \toola, a student provides a \emph{question stem} (a short section of text that describes the problem to be solved), a set of possible \emph{answer options} (between two and five alternatives are allowed, exactly one of which must be selected as the correct answer) and an \emph{explanation} (detailing the answer to the question and optionally explaining why certain alternative answers are incorrect).  Figure \ref{fig:peerwise_question_example} illustrates an example of the stem, options and explanation for a question published by a student in an introductory MATLAB programming course.  

\begin {figure}[h]
\centering
\includegraphics[width=0.95\textwidth]{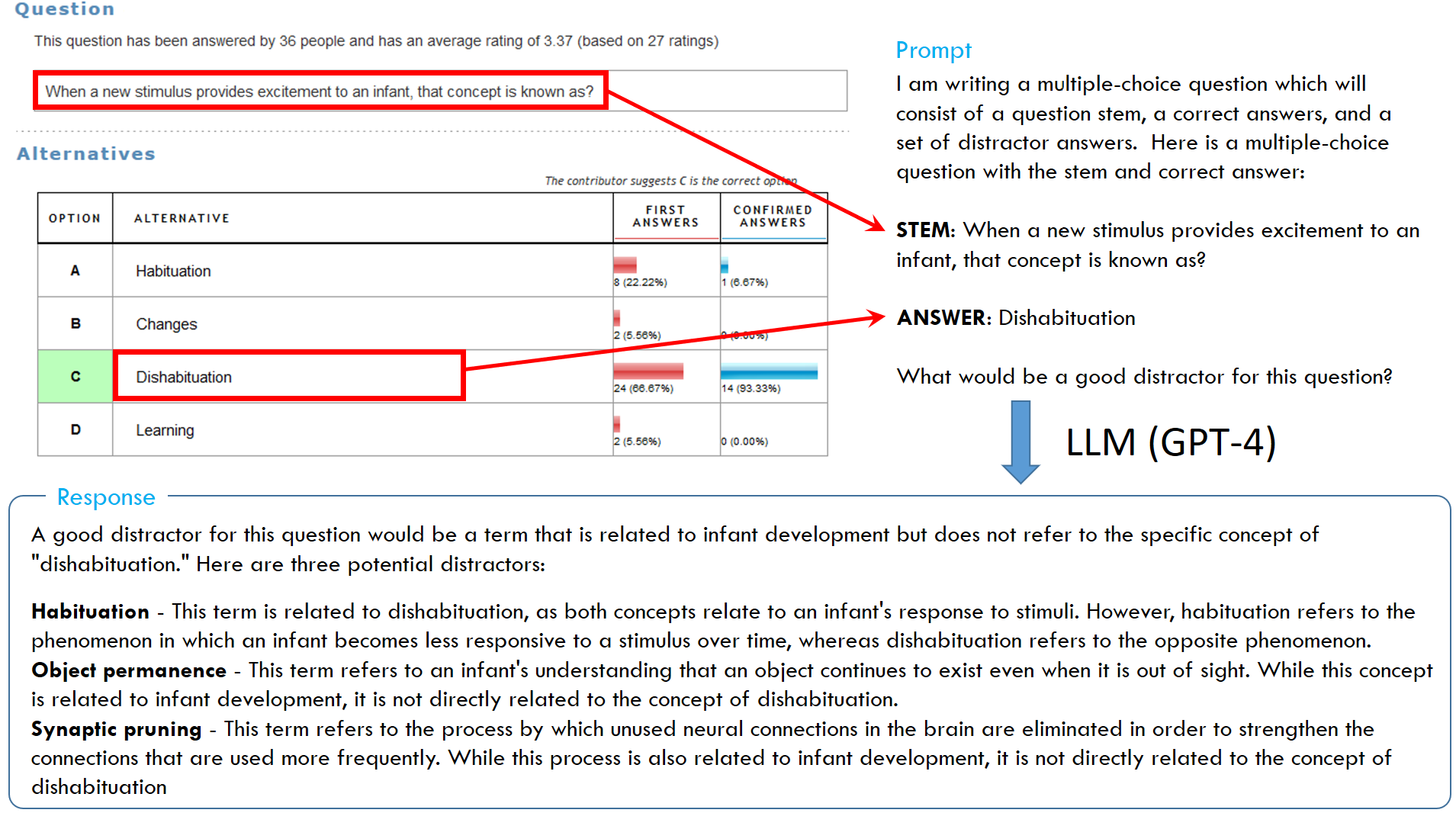}
\caption{Example of using a large language model to aid the creation process \label{fig:gpt4}} 	
\end {figure}

\begin {figure}[h]
\centering
\includegraphics[width=0.7\textwidth]{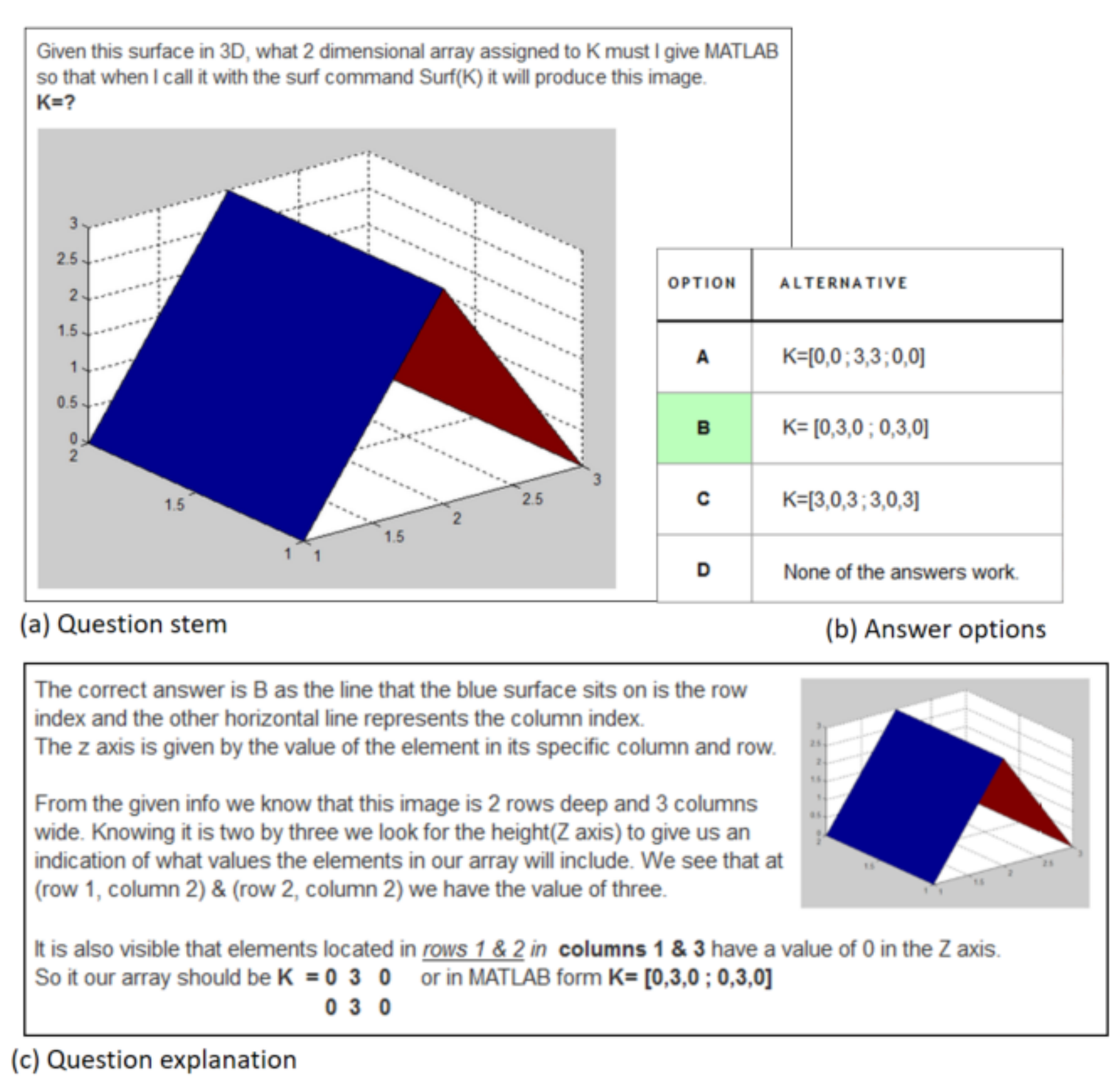}
\caption{ Example of question stem, options and explanation \label{fig:peerwise_question_example}} 	
\end {figure}  

As soon as a student publishes a question to the course repository, it is available to all other students in the course to answer and evaluate.  That is, there is no formal validation or evaluation step prior to the question being visible to other students.  To assist students in finding questions of interest during utilisation of large question banks for practice, question authors are able to assign topic ``tags'' to their questions which can be used for searching the question bank. 
{
One obvious human-AI partnership that can enhance the ``create'' activity in a learnersourcing environment is the use of a large language model, such as GPT-4, to provide automatic guidance to an author generating content.  Consider, for example, a student using \toola to create a multiple-choice question as shown in the upper left portion of Figure \ref{fig:gpt4}.  As soon as the author of this question has crafted the question stem and indicated the correct answer, these two elements can be incorporated into a simple prompt (shown on the upper right of Figure \ref{fig:gpt4}) and sent automatically to a large language model (in this example, GPT-4 is used).  The response from the model, shown in the bottom portion of Figure \ref{fig:gpt4}, presents three possible distractors for this question.  These can then be presented to the student to help them craft an effective question.  In this particular example, the question was already published and the responses selected by students utilising the question are illustrated by the red and blue histograms in the ``First answers'' and ``Confirmed answers'' columns, but naturally this data would not be available to the author at the time of writing the question.  Note that the first distractor suggested by the language model, ``habituation'', actually turns out to be the most effective distractor in practice (i.e. the incorrect option selected by more than 20\% of students attempting this question).  The fact that this option was not provided to the language model as part of the prompt indicates that it is able to formulate suggestions that have utility.  The other suggested distractors may have proven more effective, had they been used, than the current distractors which were rarely selected. 
}

\subsubsection{Evaluate in \toola}

Any student may evaluate any of the questions in the question bank, however they must first attempt the question by submitting an answer.  Submitted answers are assessed through comparison with the question author's suggested answer, and the answers submitted to the question by other students. \toola generates one of seven possible feedback responses each time an answer is submitted.  For example, if the submitted answer matches the author's suggested answer and that is also the most popular answer selected by other students, then the feedback response includes a solid green tick and a short descriptor message.  An answer is deemed incorrect, and denoted by a solid red cross, if that answer differs from the author's suggested answer yet the suggested answer is the most popular answer selected by other students.  Variations to this feedback, with appropriate descriptors, are shown when there is mixed agreement between the submitted, author's and most popular answer.  

After submitting an answer and receiving this feedback, students are also shown the question author's explanation and a histogram of the options selected by other students.  At this stage, a student can add a comment to the comment thread for the question, and they can evaluate the question by submitting a rating for its difficulty (3-point scale) and quality (6-point scale).  There is also an option to ``flag'' any question deemed inappropriate, which will bring it to the attention of the course instructor facilitating their oversight role.  Finally, in light of seeing the feedback and discussion on the question, the student can optionally submit a ``confirmed'' answer which indicates the option they believe is correct.  Alongside the ratings, response histograms and comments, these confirmed answers serve to improve the quality of the feedback students receive when utilizing the resource for practice.  Figure \ref{fig:peerwise_rating_comment_example} shows an example of the answer histogram for a question in a cellular biology course (with initial question attempts shown in red, and ``confirmed'' answers in blue), and a short excerpt from the comment thread to a question in a computer architecture course. 

\begin {figure}[h]
\centering
\includegraphics[width=0.7\textwidth]{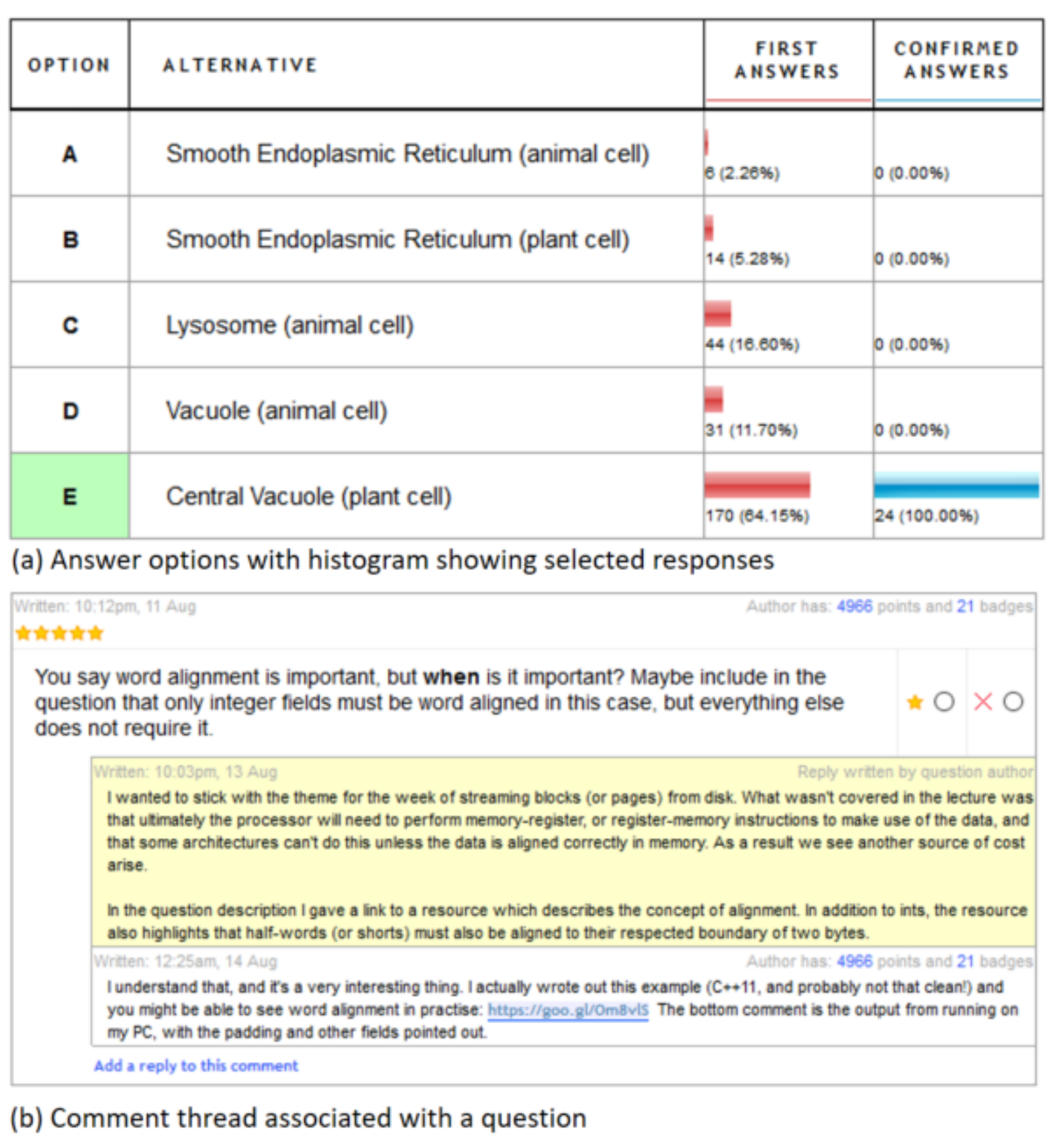}
\caption{ Answer histogram showing distribution of selected responses and comment thread \label{fig:peerwise_rating_comment_example}} 	
\end {figure}

\subsubsection{Utilise in \toola}
The prior evaluation step provides two mechanisms for helping students locate high quality questions when utilising the repository for practice.  Firstly, the difficulty and quality ratings are aggregated and can be used directly by a student to avoid low quality questions and to select questions at a suitable level of difficulty.  Secondly, after evaluating a question a student can choose to ``follow'' the question author if they find the question was particularly helpful.  Consistent with the anonymous interface design, students do not know the identities of the authors when choosing who to follow, and instead must make quality judgements on the basis of the content.  Once following a particular author, a student gains access to all of the other questions that author has created in a separate section of the tool.  As a way of incentivising the authoring of useful questions, students can see how many of their peers are following them.  

A common use of question repositories in \toola is for review and practice purposes leading up to summative tests and exams.  Prior work has shown that answering activity in \toola typically increases rapidly before a test \citep{denny2015generating}, and that answering questions is strongly predictive of subsequent test performance \citep{denny2018empirical, snow2018discursive}.  Instructors are also able to make use of the questions, for example by reviewing a question repository to identify topics that are challenging for students, or by selecting high-quality questions for use on summative tests and exams.  For example, \cite{huang2021selecting} showed that with some basic coaching, students using \toola were able to produce many questions that performed just as well when used on high-stakes exams as questions authored by academics.

\subsubsection{Oversight in \toola.} \toola was originally inspired by the contributing student approach described by \citep{hamer2006some}, and thus takes the view that students are primarily responsible for both producing and moderating the resources.  As such, instructor oversight of the learnersourcing process is fairly minimal.  

To provide some initial structure for students when authoring questions, instructors can define a set of ``course tags'' which are topics that are shown to students during the authoring step at the point they are prompted to tag their question.  Defining these tags can help to minimise fragmentation of topics.  Instructors can also post ``administrator comments'' as part of the comment thread for any question.  Such comments are highlighted as being posted by the instructor, and appear separately from student written comments.  

Students can, of course, edit their own questions but cannot edit questions written by other students.  They are limited to providing feedback via a question's comment thread or flagging questions, at the time of rating, that they deem are inappropriate.  Instructors have the ability to edit or delete questions in the repository, and can review questions which have low quality ratings or which have been flagged by students.

\subsection{\toolb}\label{sec:RiPPLE}
\begin {figure*}[h]
\centering
\includegraphics[width=0.5\textwidth]{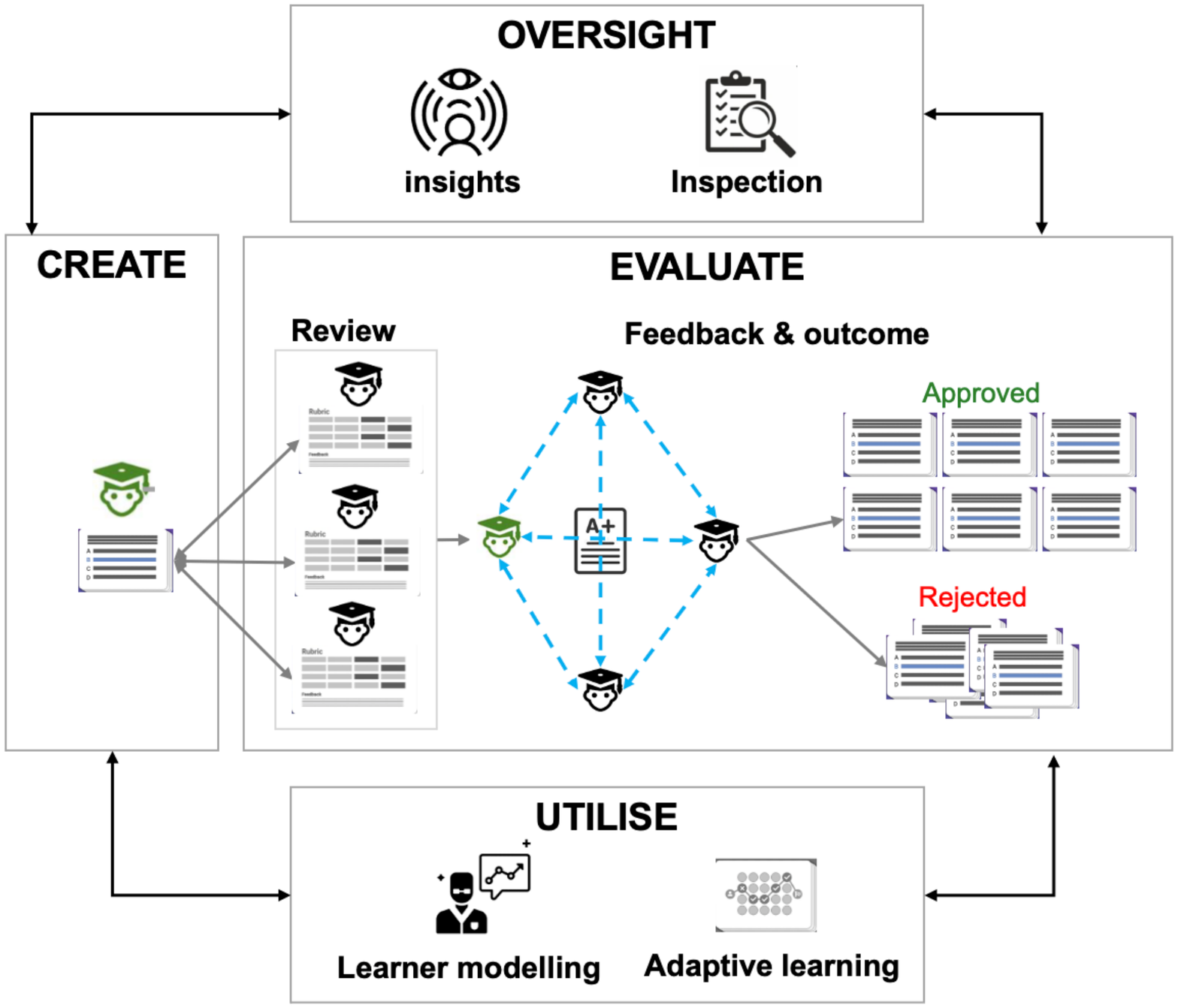}
\caption{ Overview of \toolb.  \label{fig:rippleframework}}	
\end {figure*}  
A full description of \toolb is provided in \citep{khosravi2019ripple}. 
Here, we provide a brief description based on our proposed learnersourcing framework presented in Figure~\ref{fig:framework}. At its core, \toolb is an adaptive educational system that relies on learnersourcing for content creation. \toolb can be used as a standalone system or be integrated into many popular LMSs using the Learning Tools Interoperability (LTI) standard. \toolb supports two types of roles: instructors and students. To use the system in a course, an instructor creates a \toolb offering and adds a set of topics, and optionally imports resources from other \toolb offerings. This enables instructors to import resources from their past offerings as well as share resources with other instructors with or outside their institution. In \toolb, students own the intellectual property rights of any content that they create and are free to share their content with others as they desire. However, to enable students to benefit from each other’s contributions, the terms and conditions of using the platform request students to provide a non-exclusive licence to host, use, distribute, modify, run, copy and publicly display their content.

Since 2018, \toolb has been used in over 150 course offerings with over 30k students who have created over 80k resources which have received over 300k peer reviews. Figure~\ref{fig:rippleframework}  provides an overview of the operation of \toolb with respect to our framework. We elaborate on each part of the framework in the following sections.

\subsubsection{Create in \toolb}
\begin {figure*}[h]
\centering
\includegraphics[width=1\textwidth]{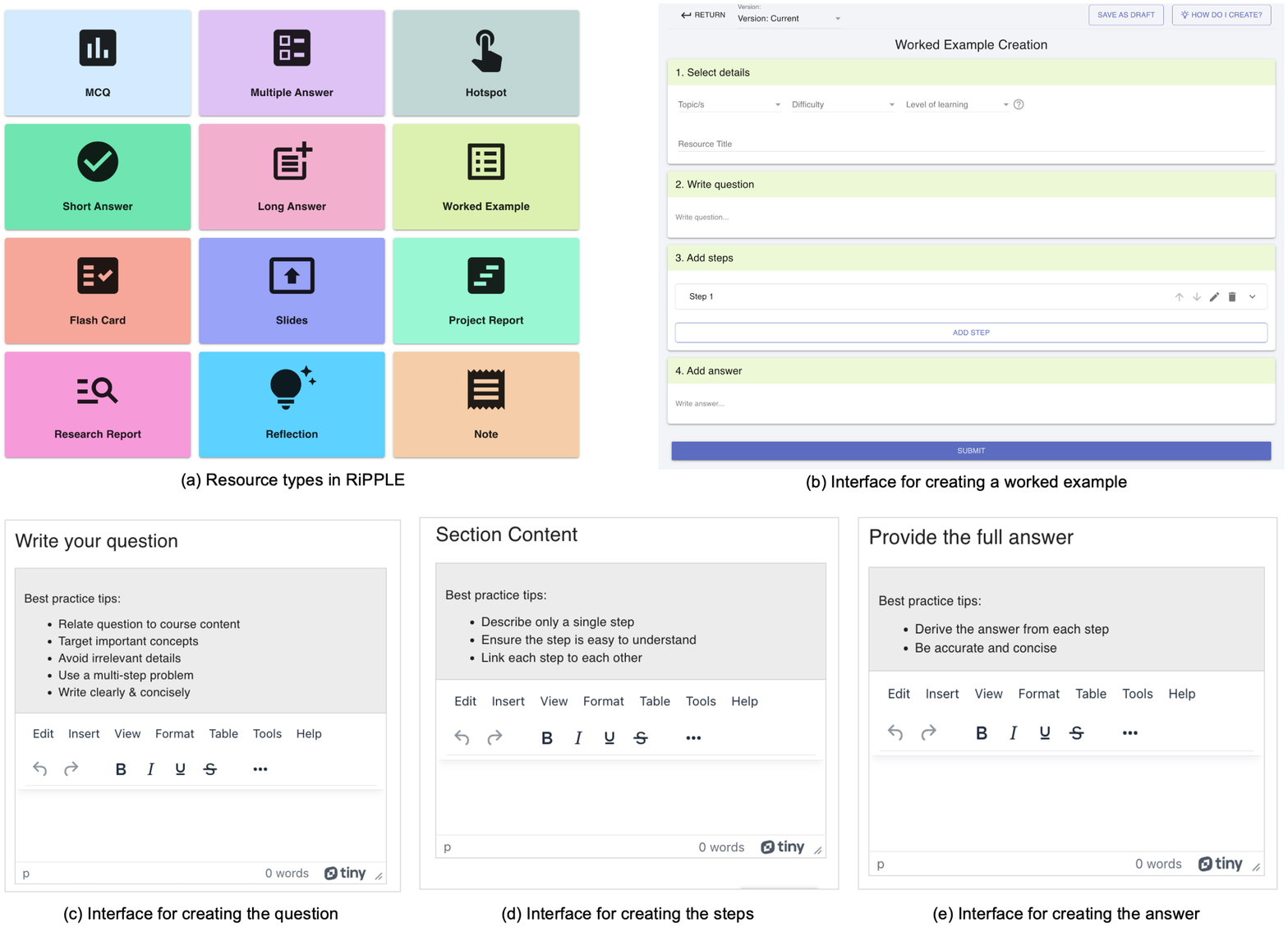}
\caption{ Creation in \toolb.  \label{fig:ripplecreate}}
\end {figure*}
Both students and instructors can create learning resources in \toolb. Originally, similar to many other learnersourcing systems, \toolb only supported the creation of MCQs. However, during the last few years, as demonstrated by Figure~\ref{fig:ripplecreate}(a), we have added support for the creation of various other educational resource types such as flashcards, hotspots, worked examples as well as short and long answer questions. Based on popular demand from instructors, we have also included resource types such as reflections, research reports and project reports, which allow students to share original contributions that are not directly related to covering content from a course curriculum. Figure~\ref{fig:ripplecreate}(b) illustrates the interface used for creating a worked example. Students are asked to identify the topic(s) associated with the resource, its level of difficulty, and it's corresponding position within Bloom's revised taxonomy of learning objectives \citep{krathwohl2002revision}. Given that we expect not all students to be familiar with this taxonomy, the platform provides a description and an example resource for each level of learning. Figure~\ref{fig:ripplecreate}(c)-(e) show the interfaces and the provided best practice tips for writing the question, steps and a sample solution.   

\subsubsection{Evaluate in \toolb}
Resources created by students go through a formal peer-review process \citep{darvishi2021employing}. Upon availability of a student to peer review a resource (i.e., the student goes on the moderation tab on the platform), \toolb selects and presents a non-evaluated resource to the student. Each resource type has an associated rubric for evaluation, and these rubrics share a similar structure. All rubrics have a set of criteria, a statement capturing the evaluators' perceptions of the overall quality of the resource, a statement capturing their confidence in their judgement and written feedback to justify their decisions. However, specific details the underlying criteria for evaluating various resource types have changed over time. The initial version of the rubric, shown in Figure~\ref{fig:rippleevaluate}(a), used a Likert scale to capture students' responses to criteria (alignment with course content, correctness and coherence of the resource), decision and confidence level. Analysis of more than 40,000 student evaluations based on this rubric revealed that it led students to lenient marking as over half of the students provided the highest quality rating to the resources. 
\footnote{Informed consent from students and approval from the University of Queensland was received for reporting aggregated stats on human participants' engagement.}


We also noticed that the average length of the provided comments was only eight words, which meant very little support was provided for their judgement. To address some of these shortcomings, we updated the rubric to what is shown in Figure~\ref{fig:rippleevaluate}(b). This rubric included additional criteria that referred to the appropriate level of difficulty and critical thinking. In addition, we moved away from Likert scale statements, which are commonly used to capture perceptions in surveys, to words that refer to the quality of outcome ranging from poor to outstanding, which is more commonly used in rubrics. Finally, the updated rubric specifically asked students to justify their decision and provide feedback rather than just having space for a comment without specific instructions. Analysis of more than 30,000 student evaluations based on the updated rubric showed a significant shift in their responses where the most common rating response moved from the highest rating (5) to the second highest rating (4). The mean length of the associated comments supporting the quality ratings also increased to 13 words. 

Figure~\ref{fig:rippleevaluate}(c) illustrates the next main update to the rubric, described in \citep{darvishi2022incorporating}, which aimed to improve the quality of the provided feedback. Informed by higher education research, we built a set of training materials (accessible by clicking on the ? button next to where they justify their response) and a self-monitoring checklist for students to consider while writing their reviews. We also developed natural language processing-based quality control functions that automatically assess feedback submitted and prompt students to improve, if necessary.  Analysis of over 190,000 student evaluations based on this rubric indicated that we were successful in almost doubling the length of the provided feedback from an average of 13 to 25 words. 

\begin {figure}[t]
\centering
\includegraphics[width=1\textwidth]{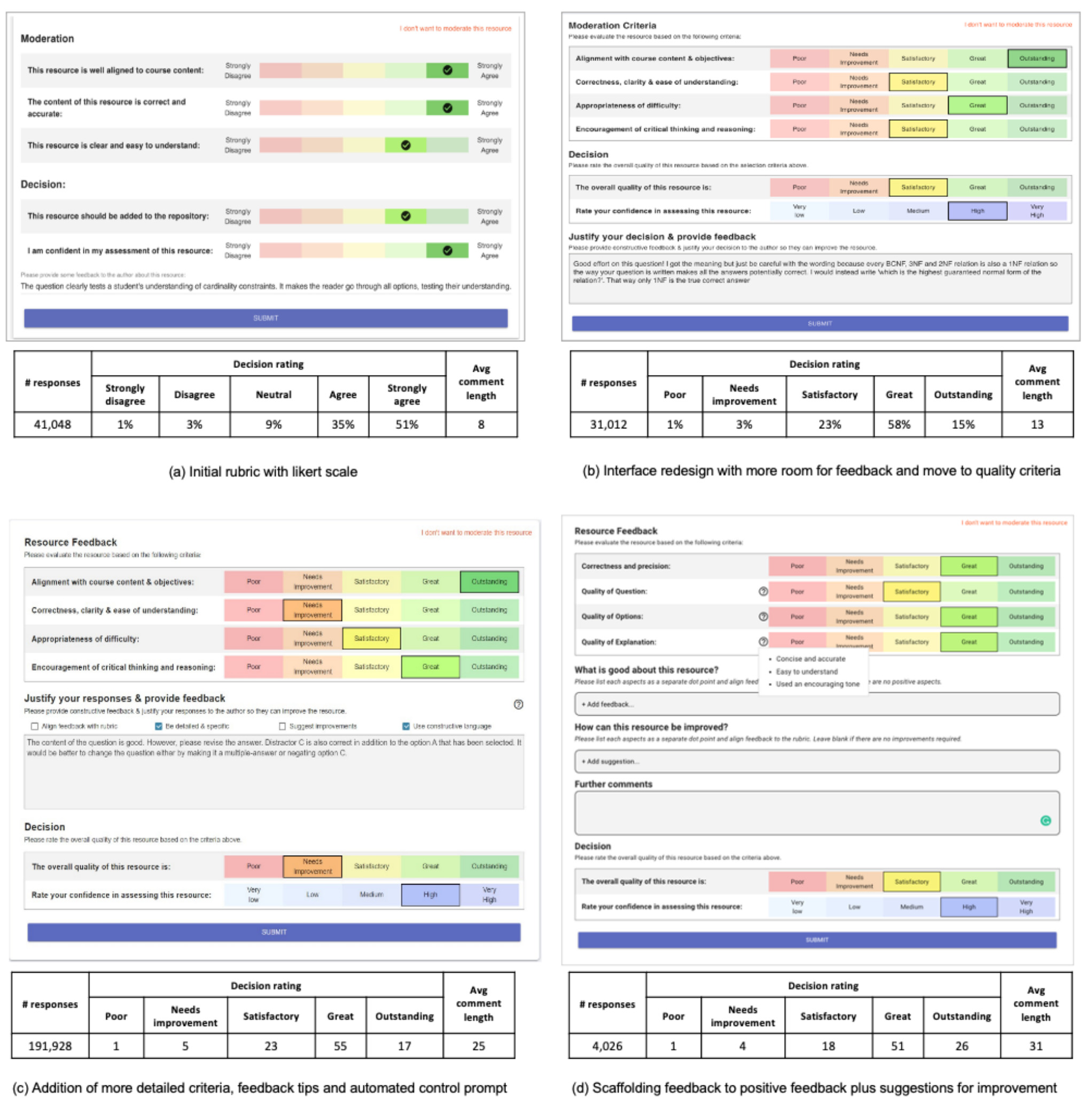}
\caption{ Evaluate in \toolb.  
\label{fig:rippleevaluate}}
\end {figure}

Figure~\ref{fig:rippleevaluate} (d) shows the current version of the rubric on the platform which has recently been deployed. This incorporates two new changes. The first change is that each criterion can now be accompanied by a list of items for the reviewer to consider in their evaluation. This change allows the platform or instructors to provide  more elaborate guidelines for a reviewer's consideration. The second change has introduced scaffolding to the feedback part where instead of writing one block of feedback students now provide a list of positive aspects about the resources followed by a list of suggestions. They can then provide further comments, if necessary.  This change was introduced as the feedback from many reviewers was generic in nature and did not include constructive suggestions on how the resource can be improved. Analysis of the impact of this new rubric is underway. Early results based on over 4,000 responses are included in Figure~\ref{fig:rippleevaluate} (d).

 A feedback and evaluation outcome interface (as shown in Figure~\ref{fig:ripplefeedback}), shares the results with the author and the reviewers, asking them to vote on the helpfulness of the evaluations and determine whether or not they agree with the outcome (approved or denied). We have collected ore than 30,000 responses since the interface was added to the platform. It is encouraging to see that students generally trust the system; only 2\% of the responses disagreed with the outcomes of the peer assessment process and fewer than 4\% mentioned they were unsure (see \citep{darvishi2022trustworthy} for more details).
 
 \begin {figure}[t]
\centering
\includegraphics[width=0.6\textwidth]{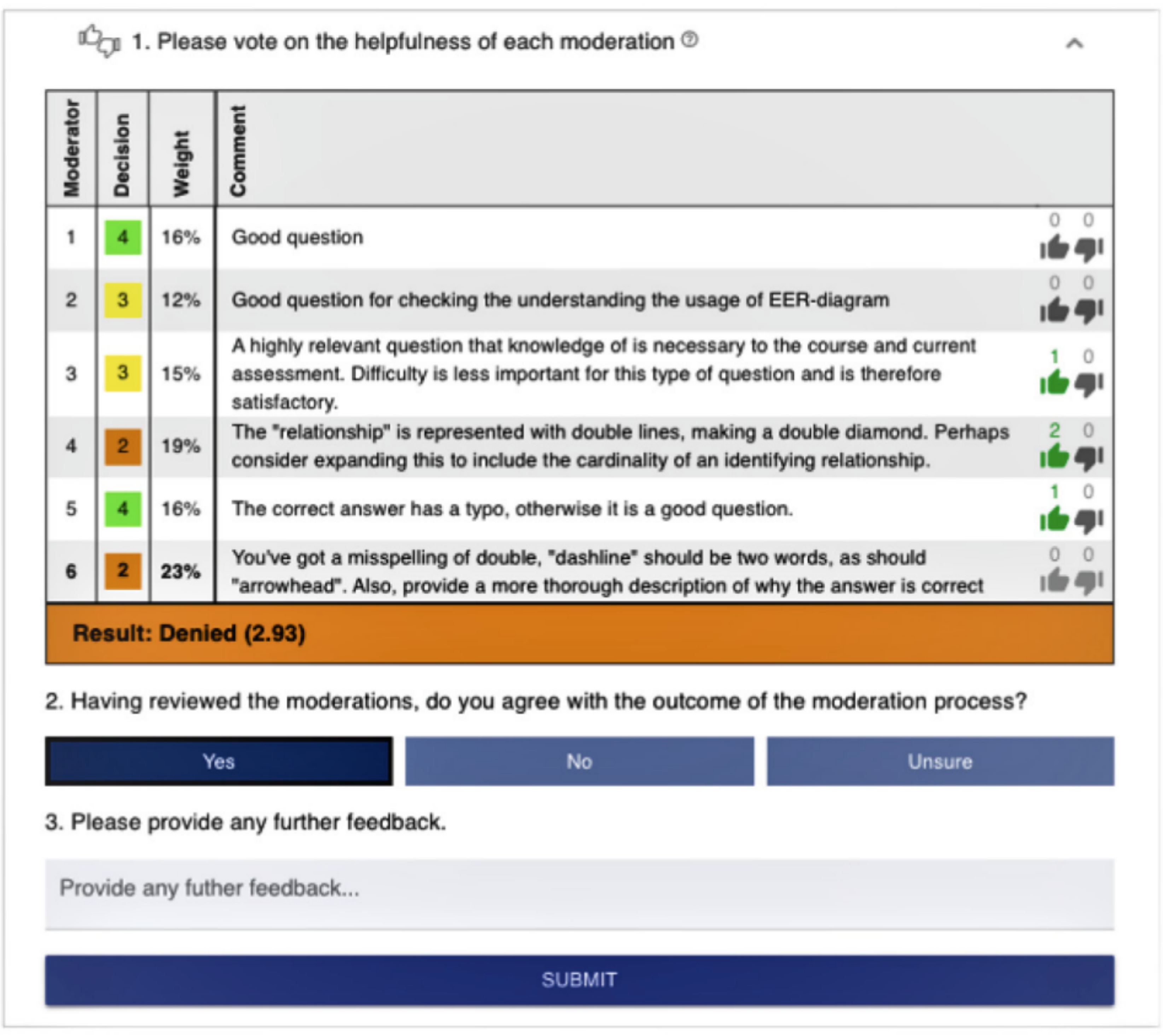}
\caption{ Feedback in \toolb.  
\label{fig:ripplefeedback}}
\end {figure}
\begin {figure*}[h]
\centering
\includegraphics[width=0.8\textwidth]{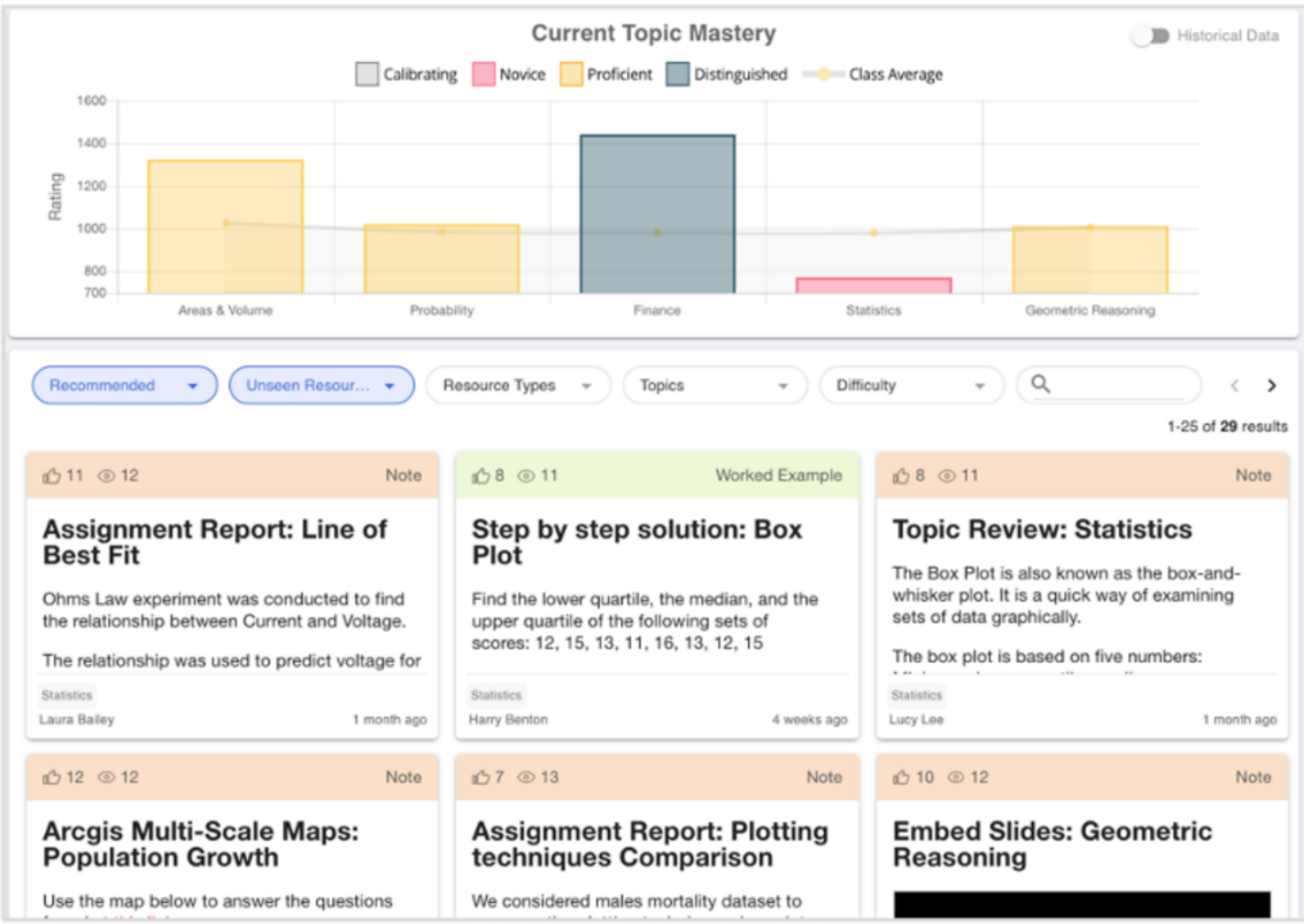}
\caption{ The open learner model and recommendation interface in \toolb.  
\label{fig:ripplepractice}}
\end {figure*}

\subsubsection{Utilise in \toolb}
Figure~\ref{fig:ripplepractice} illustrates the interface used for providing personalised practice opportunities for students. The top part of the figure represents an interactive visualisation widget, in the form of an open learner model \citep{Bull2020,abdi2021open}, that allows students to view an abstract representation of their knowledge state based on a set of topics associated with a course offering. The colour of the bars, determined by the underlying algorithm modelling the student, categorises competence into three levels. Namely, for a particular unit of knowledge, red, yellow and green signify inadequate competence, adequate competence with room for improvement, and mastery, respectively. Currently, \toolb employs an Elo-based rating system for approximating the knowledge state of users with the results translated into coloured bars \citep{abdi2020modelling}. The lower part of the screen displays learning content from the repository of approved resources that are recommended to a student based on their learning needs using the recommender system outlined in \citep{khosravi2017}. At a high level, this system recommends easier content on topics where students are developing mastery and harder content on topics for which students have already developed mastery. Students also have the ability to search for resources based on various criteria such as the resource type, topics and difficulty.

\subsubsection{Oversight in \toolb}
One of the main design guidelines of \toolb is to ensure that it optimally uses the minimal availability of instructors. To do so, \toolb has an instructor landing page, shown in Figure~\ref{fig:rippleinsights}. The top part of the page displays statistics based on overall use  and use from the previous week on study sessions as well as resources created, evaluated and answered. It then provides information about the completion status for the latest round of assessment and weekly highlights about students' achievements and popular resources. One of the main ways we have tried to optimally use instructor time is to add suggested actions. \toolb currently provides five types of suggested actions including inspecting resources that benefit the most from academic judgement, reviewing evaluations that are flagged as ineffective, submitting assessment grades,  nudging at-risk students (e.g., those who have not logged in or completed assessment) and congratulating high achievers on their achievements.  The bottom of the page includes an analytical toolbox that provides answers to a list of questions in relation to students' performance and engagement. For each of the questions instructors can set the start and end date for data being reported to view class-level or individual-level trends using various visualisation types (e.g., bar charts, box plots). 

\begin {figure*}[h]
\centering
\includegraphics[width=0.8\textwidth]{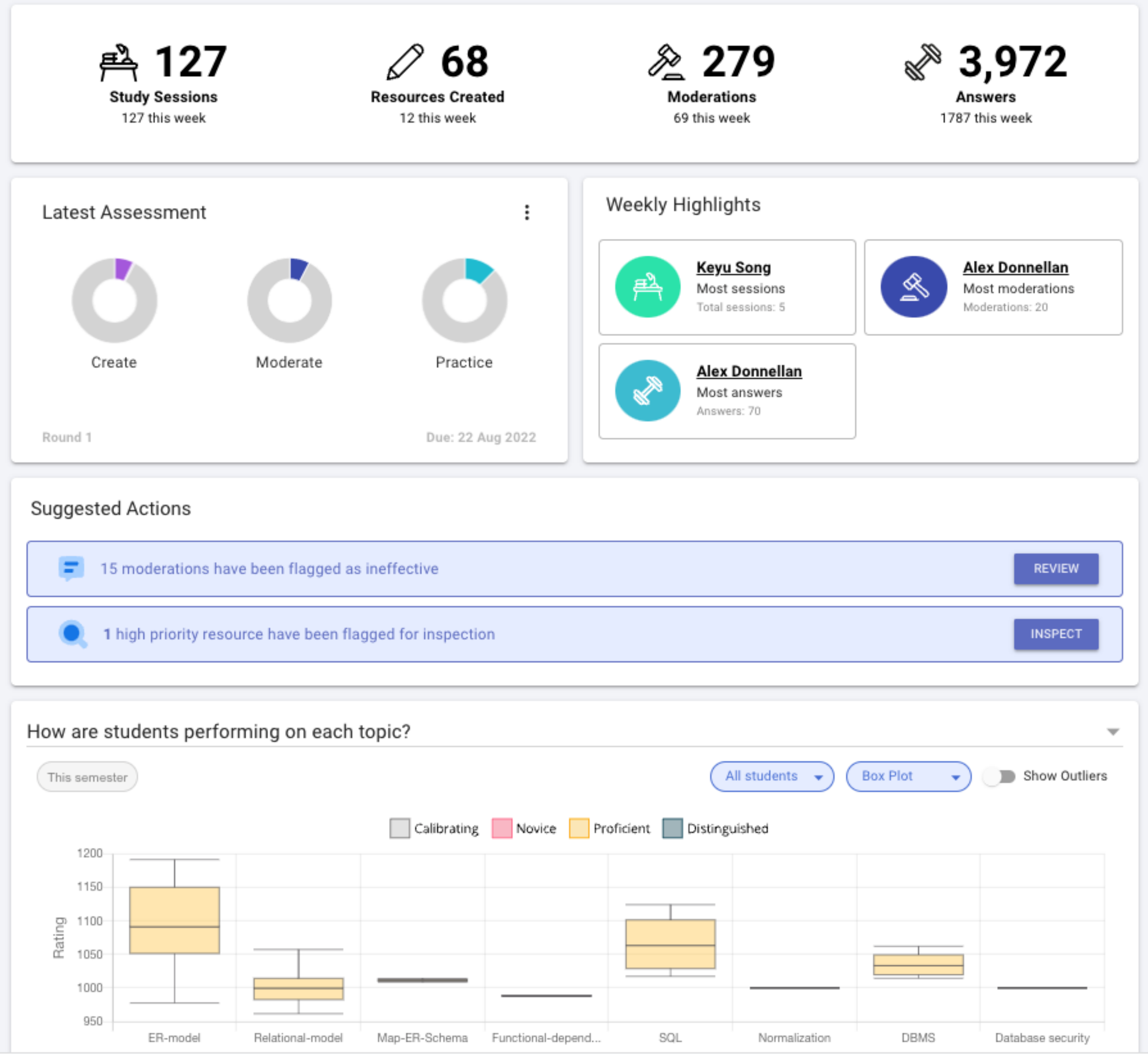}
\caption{ Weekly insights in RiPPLE  
\label{fig:rippleinsights}}
\end {figure*}
\begin {figure*}[h]
\centering
\includegraphics[width=01\textwidth]{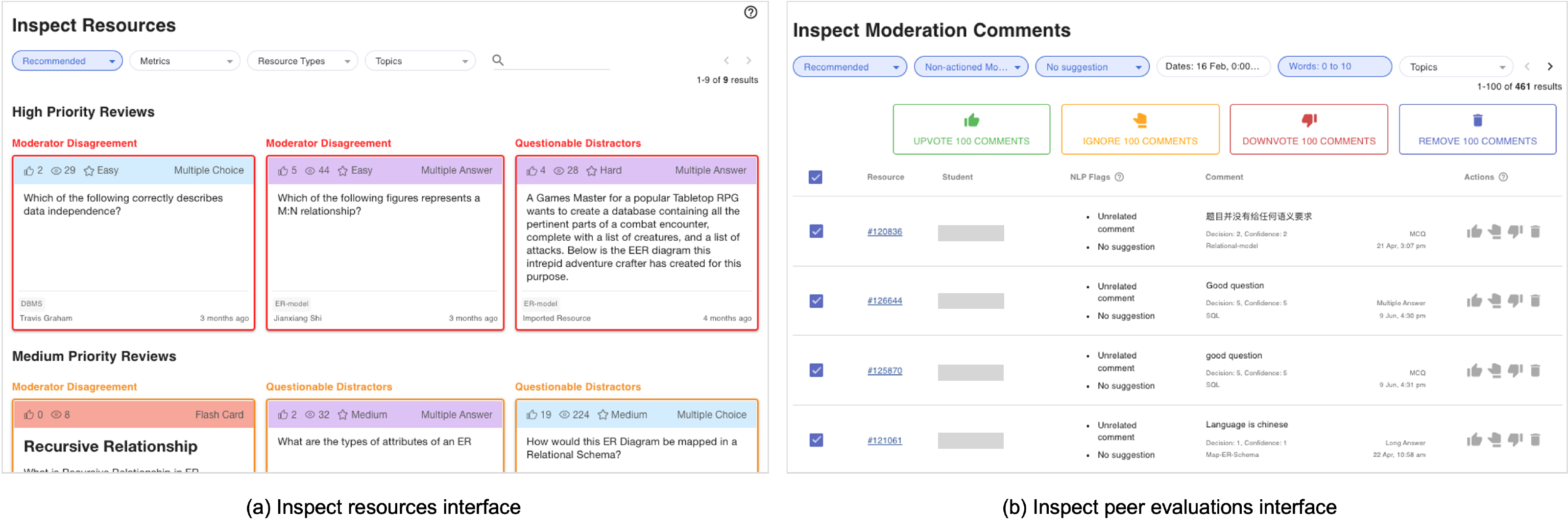}
\caption{ Resource and peer review inspection in \toolb. The name of students has redacted in the figure.  
\label{fig:rippleinspect}}
\end {figure*}

Figure~\ref{fig:rippleinspect} shows the underlying interface for two of the main suggested actions, namely inspecting resources and reviewing evaluations. Figure~\ref{fig:rippleinspect}(a) shows the \toolb interface for inspecting resources that have been approved but are likely to be incorrect or ineffective. At a high level, it employs a range of human-driven metrics (e.g., high disagreement in reviews, a high ratio of downvotes in comparison to upvotes) and data-driven metrics (e.g., assessment items that have a low discrimination index or questionable distractors where the popular answer is not the one proposed by the author) to categorise resources into having high, medium, low or no priority for being reviewed. \toolb uses absolute and relative points of comparison to help instructors make sense of why a resource has been flagged for review (e.g., `` disagreements between student ratings for this resource is 2.8 times higher than average of the course''). 

Figure~\ref{fig:rippleinspect}(b) shows the \toolb interface for searching the reviews.  It enables instructors to set a date range to identify reviews with a particular word count range on a particular set of topics. In addition, we use the functions outlined in \citep{darvishi2022incorporating} to enable instructors to identify whether or not reviews include a suggestion. Once instructors select a set of reviews, they can apply one of the following actions: upvote it to provide positive feedback;  ignore it so that it wouldn't show up in the search again; downvote it and provide feedback on its ineffectiveness; or finally remove the review so that it wouldn't count towards meeting assessment requirements for the reviewer.  

{
\section{Challenges and Implications for Practice and Research}
The adoption of our learnersourcing framework may lead to certain considerations that need to be addressed with care. This section elaborates on these challenges and implications, and also pinpoints areas that call for further exploration and research.

 The process of educating students to become proficient creators and evaluators of content, along with the duty of content creation itself, are tasks that demand considerable time. Therefore educators that are considering taking on learnersourcing may need to think about how these responsibilities can be balanced against other academic commitments in their course to avoid overloading students. Moreover, some students may not fully understand or appreciate the advantages of learnersourcing. Thus, incentivising them to participate may necessitate educators to explicitly discuss the benefits and rationale for its inclusion in their curriculum. Empirical studies focusing on the most effective strategies for incetivising students and training them in content creation, as well as assessing the impact of these strategies on student engagement, outcomes, and the quality of resources produced, could provide substantial insights for effective adoption of learnersourcing.

The utilisation of AI assistance to aid students in content creation and evaluation is a potentially beneficial and promising strategy. However, if employed without care, it may lead to overdependence, where students might excessively rely on large language models to create and evaluate content on their behalf. This could result in several negative outcomes, such as content lacking originality, not aligning with course material, or simply being factually incorrect. The design, development and validation of interfaces and scaffolds incorporating large language model APIs into learnersourcing systems, to help students more effectively contribute, presents a promising field of future research.

While the frameworks include multiple strategies to optimise the utilisation of the instructor’s efforts, a certain level of monitoring is still required to identify misbehaviour such as carrying out detrimental peer reviews, plagiarising content, or creating inappropriate resources. This monitoring not only serves as a means of maintaining content and behavioral standards, but also serves as a demonstration of reliability, providing assurance to both educators and students that the system is trustworthy and dependable. In more minor cases, the created content may lack relevance or have large coverage gaps based on the course objectives. These actions could undermine students’ trust in learnersourcing and seeing its benefits. The design, development and validation of strategies that provide actionable insights and pedagogical interventions that best assist educators in facilitating learning, identifying students who require assistance and overseeing the creation and evaluation of contributions of students in learnersourcing systems, all while demonstrating their reliability, hold substantial potential for future research

Utilising models that foster content co-creation among several students might provide the dual advantages of promoting collaborative learning and enhancing the quality of content produced. However, this approach is also prone to issues such as group dynamic conflicts, free-riding, ensuring fairness in grading, and evaluating individual contributions which are inherent to group-based learning and assessment tasks. Educators may need to think about approaches for dealing with such conflicts if they are considering enabling multi-student learnersourced submissions.
}

\section{Conclusion}\label{sec:conclusion}

Widespread changes to learning and teaching alongside rapid growth in the use of digital tools in education and the vast data they collect are presenting new opportunities for the application of artificial intelligence in the classroom. One activity that appears uniquely placed to benefit is learnersourcing. Given that it is primarily student-driven, artificial models of intelligence can be employed to enable students to develop transformative competencies, such as creating new value, developing self- and co-agency skills and improving the learning experience while reducing the need for instructor expertise and oversight.

This paper introduced a framework that considers the existing learnersourcing literature, the latest insights from the learning sciences and advances in AI to offer a blueprint for developing learnersourcing systems.
The framework is presented in the form of questions that correspond to creating content, evaluating the quality of the created content, considerations for how learnersourced contributions can be utilised and the role of instructors in facilitating learning via learnersourcing. 

A common theme across all four dimensions is the
need for human-AI partnerships for advancing learnersourcing. In terms of content creation and evaluation, advances in NLP and generative models provide space for AI to play a fundamental role in the co-creation of content with humans and to assist with the automated evaluation of its quality. For utilising learnersourcing content, the use of AI can help in developing explainable recommender systems that model students' mastery and assist them in engaging with content that best suits their learning needs. Finally, which respect to oversight, AI can partner with instructors to help them optimally use their time in reviewing content that benefits the most from their judgement and assisting students that need their help. To maintain a clear connection between the discussion and the literature pertaining to each dimension of the framework, we have included discussion points within the sections where they are presented. The two presented case studies demonstrate the application of our framework in the context of two vastly different and well-adopted learnersourcing systems.

Despite a long history of exploration, the development and large-scale adoption of learnersourcing systems are still in their early stages and much more fundamental work is needed before they can achieve their full potential. We call upon the educational research and practitioner communities to review and critique our framework and to contribute to advancing the field through the development of scientifically grounded and empirically validated systems that can help in accelerating the development and adoption of learnersourcing. In relation to human-AI partnership for creating and evaluating content, we highlight many opportunities, but there is still much that needs to be done for this vision to reach its full potential. On the educational side, there is a need for preparing students and instructors for collaboration with AI, which requires large-scale upskilling and training programs in data and digital literacy. There is also a need for conducting educational research to determine the effect size of various human-AI partnerships in creating content on student learning. Relative to social science, there is a need for developing new policies that assign ownership of content and copyright among human and AI collaborators in a fair manner. On the technical side, more effective methods of assisting humans in the create, evaluate, utilise and oversight processes are required. In particular, development of discipline-focused large language models may strengthen the quality of novel educational content AI can create. Finally, in relation to ethics, there is a need for exploring spot-checking and human-in-the-loop models that fairly treat all students and gives them equal opportunity for learning and receiving feedback.

\bibliography{refs}
\bibliographystyle{cas-model2-names}

\end{document}